\documentclass[%
reprint,
superscriptaddress,
amsmath,amssymb,
aps,
pra,
]{revtex4-2}

\usepackage{graphicx}
\usepackage{dcolumn}
\usepackage{bm}
\usepackage{color}
\usepackage{soul}
\usepackage{rotating}
\usepackage{amsmath}
\usepackage{mathrsfs}

\begin{document}
\title{Floquet quantum multiparameter estimation with periodic-driving-induced topological phase transition}

	\author{Yu Yang}
	\email{yangyu1229@hotmail.com}
	\affiliation{Ministry of Education Key Laboratory for Nonequilibrium Synthesis and Modulation of Condensed Matter, Shaanxi Province Key Laboratory of Quantum Information and Quantum Optoelectronic Devices, School of Physics, Xi’an Jiaotong University, Xi’an 710049, China}
	
	\author{Yuyang Tang}
	\affiliation{Ministry of Education Key Laboratory for Nonequilibrium Synthesis and Modulation of Condensed Matter, Shaanxi Province Key Laboratory of Quantum Information and Quantum Optoelectronic Devices, School of Physics, Xi’an Jiaotong University, Xi’an 710049, China}
	
	\author{Pei Zhang}
	\affiliation{Ministry of Education Key Laboratory for Nonequilibrium Synthesis and Modulation of Condensed Matter, Shaanxi Province Key Laboratory of Quantum Information and Quantum Optoelectronic Devices, School of Physics, Xi’an Jiaotong University, Xi’an 710049, China}
	
	\author{Fuli Li}
	\email{flli@xjtu.edu.cn}
	\affiliation{Ministry of Education Key Laboratory for Nonequilibrium Synthesis and Modulation of Condensed Matter, Shaanxi Province Key Laboratory of Quantum Information and Quantum Optoelectronic Devices, School of Physics, Xi’an Jiaotong University, Xi’an 710049, China}

\begin{abstract}
Periodically driven systems provide a powerful platform for quantum multiparameter estimation.
Constructing a static effective Hamiltonian in a proper rotating frame is commonly employed to assess the attainable precision.
However, such an approach becomes nonfeasible for more general time-periodically driven systems.
To tackle this dilemma, we develop a quantum multiparameter estimation strategy in the Floquet theory framework.
The contributions of Floquet eigenmodes, quasienergies, and multi-photon processes to the quantum Fisher information matrix and measurement incompatibility are determined, respectively.
Moreover, this approach is applied to a ring-shaped Rashba spin-orbit interferometer model exhibiting the topological phase transition (TPT).
In the vicinity of the TPT boundary, we reveal a pronounced enhancement in the estimation precision of multiple parameters with the Heisenberg limit scaling and even higher. 
Meanwhile, the measurement incompatibility vanishes in an oscillatory manner, and the stroboscopic projective measurement enables the highest estimation precision achievable.
This work provides a complete Floquet picture for time-dependent critical quantum multiparameter estimation.
\end{abstract}
\maketitle

\section{Introduction}\label{Sec:Introduction}
The time-periodic driving plays a crucial role in quantum simulation~\cite{Floquetinsulator,Floquetphase,NC2025}, which is utilized for exploring exclusive phases of matter within the intricate dynamics evolution. 
In quantum metrology, the time-periodically driven systems are also comparably significant.
Notable applications include the development of magnetometry using dynamical decoupling sequences~\cite{MP,Lang2015,Enhanced}, the investigation of noise effects on sensing precision in open quantum systems~\cite{An2023,NM,drive,driveNew}, and so on.
These works unavoidably involve the problem of time-dependent single-parameter or multiparameter estimation.
Differing from the time-independent quantum metrology~\cite{V2006,Paris2009,Control,Multi2020}, the complexity embodied in the time evolution operator makes the estimation precision of parameters difficult to be obtained analytically.
The method of constructing a static effective Hamiltonian in a proper rotating frame is commonly employed to assess the attainable precision~\cite{Pang2017,Hou2021}.
However, this static Hamiltonian does not always exist for the more general time-periodically driven system. For instance, in equation~(\ref{eq:H}) the time-dependence cannot be removed by the rotation transformation. 
Moreover, the realistic sensing tasks generally involve estimating multiple unknown parameters.
How to perform quantum multiparameter estimation for a non-static effective Hamiltonian remains an open question.

The Floquet theory~\cite{Shirley1965,Schmidt1992,Chu2004} gives a pathway of transferring the time-dependent Hamiltonian into the time-independent and infinite-dimensional effective Floquet Hamiltonian in an extended Hilbert space (i.e., Sambe space). 
The evolution of dynamics can be numerically investigated within this space by fully diagonalizing the Floquet Hamiltonian with adequate matrix truncation.
Recent studies have demonstrated the applicability of the Floquet theory to diverse sensing issues. 
For example, the noisy quantum metrology with the Floquet bound state~\cite{An2023},
the steady-state sensing in partially accessible many-body systems~\cite{Ba2021}, 
the high-precision off-resonant microwave sensing~\cite{Lu2025}, 
and the Floquet time crystals as AC field sensors~\cite{Timecrystal}, and so on.
Another well-developed application is the Floquet engineering~\cite{Goldman2014,FloquetEng,FloquetEngNew} that relies on the time-periodic modulation of a Hamiltonian to design the quasienergy band structure for exploring the novel phases of matter~\cite{Floquetinsulator,Floquetphase,Eckardt2017}.

In quantum multiparameter estimation, the long-standing goal is to simultaneously estimate multiple parameters with the individual highest precision~\cite{Multi2020}.
However, this goal is often impeded by inherent incompatibilities from the probe state~\cite{IM3,Jarzyna2016} and the measurement scheme ~\cite{PRX2021,IM1,IM2,IM4,Vim}.
The consumed metrological resource is generally quantified by the number of probes $N$ or the evolution time $t$.
In classical sensing scenarios, the estimation precision is fundamentally limited by the standard quantum limit (SQL), which scales as $\sim N$ or $\sim t$.
By introducing entanglement~\cite{Entanglement}, squeezing~\cite{Squeezing}, coherent superposition~\cite{ICO}, and quantum criticality~\cite{FRe,F1,Critical,Arxiv}, the SQL can be surpassed, leading to the Heisenberg limit (HL) scaling $\sim N^2$ or $\sim t^2$.
Distinct from canonical strategies, quantum criticality has recently emerged as a promising resource for achieving the HL scaling~\cite{MB2021,NP2022} and even the super-Heisenberg limit scaling~\cite{Boixo2007,SHL}.
By driving the system into the critical point, the information encoded into the probe state gradually accumulates to a maximal (finite or infinite) value, which is governed by the (quasi)energy gap closing~\cite{FRe,F1,Critical,Arxiv,Ba2021}.
In particular, the topological phase transition (TPT), as a hallmark of quantum criticality, has recently emerged as a promising resource for quantum sensing~\cite{Cai2024,Bayat2024,Yang2024}.
Nevertheless, the role of time-dependent Hamiltonians in these related studies has not been systematically investigated.
Whether the advancement of the TPT to multiparameter estimation precision survives in general periodically driven systems thus remains largely unexplored.

In this work, we develop a Floquet quantum multiparameter estimation strategy for a general time-periodically driven system.
By leveraging the Floquet theory to remove the time-dependence of the target Hamiltonian, the contributions of Floquet eigenmodes, quasienergies, and multi-photon processes to estimation precision are determined, respectively.
Moreover, we apply this approach in a ring-shaped Rashba spin-orbit interferometer that exhibits the periodic-driving-induced TPT.
In the vicinity of the TPT, we systematically investigate the quantum Fisher information matrix and the associated measurement incompatibility, and employ stroboscopic projective measurement as the measurement scheme. 
Additionally, to further validate our Floquet approach, we apply it to a paradigmatic rotating magnetic field sensing scenario.

The remainder of this paper is structured as follows. 
In section~\ref{Subsec:Floquet} we recall the fundamental Floquet theory and develop a Floquet quantum multiparameter estimation strategy in section~\ref{Subsec:Fparameter}.
In the next section~\ref{Sec:Rashba}, the TPT exhibited in the ring-shaped Rashba spin-orbit interferometer is demonstrated from two aspects: the topological invariant (winding number) in section~\ref{Subsec:winding} and the total phase signature in section~\ref{Subsec:phase}.
Then the Rashba and Zeeman magnetic fields and the monochromatic driving frequency in this model are simultaneously estimated in section~\ref{Sec.FL}.
Our Floquet approach is also applied in section~\ref{Sec:Rotate} to sense a rotating magnetic field. 
Finally, the conclusions and discussions are provided in section~\ref{Sec:Conclusion}.

\section{Floquet quantum multiparameter estimation strategy}\label{Sec:Floquet}
\subsection{Floquet theory}\label{Subsec:Floquet}
This section provides the complete Floquet theory, and readers familiar with it may proceed directly to section~\ref{Subsec:Fparameter}.
The Schr\"odinger equation for a time-periodic quantum system is
\begin{eqnarray}\label{eq:SH}
	\mathcal{H}(t) |\Psi(t)\rangle=0\;,
\end{eqnarray}
with 
\begin{eqnarray}
	\mathcal{H}(t)=H(t)-i \hbar{\partial}/{\partial t}\;,
\end{eqnarray}
where the Hamiltonian satisfies $H(t)=H(t+T)$ with the period $T=2\pi/\omega$ ($\omega$ is the frequency of the monochromatic driving field).
We set $\hbar=1$ hereafter for simplicity.
One of the solutions to equation~(\ref{eq:SH}) in the Hilbert space $\mathbf{H}$ is given by
\begin{eqnarray}\label{eq:state}
	|\Psi(t)\rangle=\sum_\alpha c_\alpha e^{-i\varepsilon_\alpha t} |\Phi_\alpha(t)\rangle\;,
\end{eqnarray}
where $c_\alpha=\langle \Phi_\alpha(0)|\Psi(0)\rangle \in \mathbb{C}$, $\varepsilon_\alpha \in \mathbb{R}$, and $\{ |\Phi_\alpha(t)\rangle \}$ constitutes a complete orthonormal basis. 
$\varepsilon_\alpha$ represents the $\alpha$-th quasienergy and is defined modulo $\omega$.
$|\Phi_\alpha(t)\rangle$ denotes the $\alpha$-th quasi-stationary state and is referred to as the Floquet mode or the {\it Floquet eigenstate}~\cite{Shirley1965,Chu2004,Schmidt1992,Flo2}, which satisfies the periodicity  $|\Phi_\alpha(t)\rangle=|\Phi_\alpha(t+T)\rangle$.
By inserting equations~(\ref{eq:state}) into (\ref{eq:SH}), we get
\begin{eqnarray}\label{eq:eig}
	\mathcal{H}(t)|\Phi_\alpha(t)\rangle=H_F|\Phi_\alpha(t)\rangle=\varepsilon_\alpha |\Phi_\alpha(t)\rangle\;,
\end{eqnarray}
where the Hamiltonian ${H}_F$ is known as the {\it Floquet Hamiltonian}.
Then we explore the evolution of dynamics in an extended Sambe space $\mathbf{S}=\mathbf{H} \otimes \mathbf{T}$ where $\mathbf{T}$ is spanned by the Fourier series with the time period $T$. 
The Hamiltonian and the Floquet eigenstate are expanded with the Fourier series as
\begin{align}\label{eq:Fourier}
	H(t) &= \sum_n H^{(n)} e^{in\omega t}\;,\nonumber\\
	|\Phi_\alpha(t) \rangle &= \sum_m |\Phi_\alpha^{(m)}\rangle e^{i m \omega t}\;,
\end{align}
where $n,m \in \mathbb{Z}$.
By substituting equations~(\ref{eq:Fourier}) into (\ref{eq:eig}), we get
\begin{widetext}
\begin{eqnarray}
	&&\left(\sum_n H^{(n)} e^{in\omega t}-i {\partial}/{\partial_t}\right) \sum_m |\Phi_\alpha^{(m)}\rangle e^{im \omega t}
	=\varepsilon_\alpha \sum_m |\Phi_\alpha^{(m)}\rangle e^{im \omega t},\nonumber\\
	&&\rightarrow \sum_{nm}H^{(n)} |\Phi_\alpha^{(m)}\rangle e^{i(n+m)\omega t}+\sum_m(m\omega) |\Phi_\alpha^{(m)}\rangle e^{im\omega t}
	=\varepsilon_\alpha \sum_m  |\Phi_\alpha^{(m)}\rangle e^{im\omega t},\nonumber\\
	&&\rightarrow \sum_{km}H^{(k-m)}|\Phi_\alpha^{(m)}\rangle e^{ik\omega t}+\sum_{k} (k\omega) |\Phi_\alpha^{(k)}\rangle e^{i k\omega t}
	=\varepsilon_\alpha \sum_{k}  |\Phi_\alpha^{(k)}\rangle e^{i k\omega t}\;,
\end{eqnarray}
\end{widetext}
where $n+m=k\in \mathbb{Z}$ is utilized, and for any $k$ we have
\begin{eqnarray}\label{eq:f1}
	\sum_{m}H^{(k-m)}|\Phi_\alpha^{(m)}\rangle + k\omega |\Phi_\alpha^{(k)}\rangle=\varepsilon_\alpha   |\Phi_\alpha^{(k)}\rangle \;.
\end{eqnarray}

A complete orthonormal basis $\{|\beta \rangle\}$ in the Hilbert space $\bm H$ and another complete orthonormal basis  $\{|m\rangle\}$ in the space $\bm T$ constitute a basis in the Sambe space as
\begin{eqnarray}
	|\beta,m\rangle =|\beta\rangle \otimes |m\rangle=|\beta\rangle \otimes e^{i m \omega t}\;,
\end{eqnarray}
where $\otimes$ denotes the tensor product, $\beta$ is the N-level system index, and $m=(-\infty,\infty)$ represents the Fourier index. We then rewrite $|\Phi_\alpha^{(m)}\rangle=\sum_{\beta} |\beta\rangle \langle \beta|\Phi_\alpha^{(m)}\rangle  =\sum_\beta \Phi^{(m)}_{\alpha\beta} |\beta\rangle=\sum_{\beta,\gamma} \Phi_{\alpha\beta}^{(m)} |\gamma\rangle \langle \gamma|\beta\rangle=\sum_{\beta,\gamma}\Phi_{\alpha\beta}^{(m)}\delta_{\gamma\beta} |\gamma\rangle$ with the Kronecker delta function $\delta_{\gamma\beta}$.
Similarly,  we have $H^{(n)}|\beta\rangle=\sum_\gamma   |\gamma\rangle \langle \gamma|H^{(n)}|\beta\rangle=\sum_\gamma H_{\gamma \beta}^{(n)} |\gamma\rangle$.
Accordingly, equation~(\ref{eq:f1}) can be refomulated into
\begin{widetext}
\begin{eqnarray}
	&&\sum_{m,\beta} \Phi_{\alpha\beta}^{(m)} H^{(k-m)}  |\beta\rangle  + k\omega  \sum_\beta \Phi_{\alpha\beta}^{(k)} |\beta\rangle =\varepsilon_\alpha \sum_\beta \Phi_{\alpha\beta}^{(k)} |\beta\rangle,\nonumber\\
	&&\rightarrow  \sum_{m,\beta,\gamma} \Phi_{\alpha\beta}^{(m)} H^{(k-m)}_{\gamma \beta}|\gamma\rangle +k\omega \sum_{\beta,\gamma} \Phi_{\alpha \beta}^{(k)} \delta_{\gamma \beta}|\gamma\rangle
	=\varepsilon_\alpha \sum_{\beta,\gamma} \Phi_{\alpha\beta}^{(k)} \delta_{\gamma \beta}|\gamma\rangle,\nonumber\\
	&&\rightarrow  \sum_{m,\beta,\gamma} \Phi_{\alpha\beta}^{(m)} H^{(k-m)}_{\gamma \beta}|\gamma\rangle +k\omega \sum_{\beta,\gamma,m} \Phi_{\alpha \beta}^{(m)} \delta_{mk} \delta_{\gamma \beta}|\gamma\rangle=\varepsilon_\alpha \sum_{\beta,\gamma} \Phi_{\alpha \beta}^{(k)} \delta_{\gamma \beta}|\gamma\rangle\;.
\end{eqnarray}
\end{widetext}
Given an arbitrary $|\gamma\rangle$, we get 
\begin{eqnarray}\label{eq:HHH}
	\sum_{m,\beta}   \left(H^{(k-m)}_{\gamma \beta} +k\omega \delta_{mk} \delta_{\gamma \beta}\right)  \Phi_{\alpha \beta}^{(m)} =\varepsilon_\alpha \Phi_{\alpha\gamma}^{(k)}.
\end{eqnarray}
Equation~(\ref{eq:HHH}) is recognized as the matrix eigenvalue equation for the Floquet Hamiltonian $H_F$.
$H_F$ is an infinite-dimensional time-independent Hermitian matrix with rows identified by the indices \{$\gamma$, $k$\} and columns by \{$\beta$, $m$\}, and $\gamma (\beta)$ runs over the N-level quantum states before each change in $k (m)$.
Consequently, the dynamics evolution governed by a time-periodic Hamiltonian $H(t)$ can be mapped to the one by a time-independent infinite-dimensional Floquet Hamiltonian $H_F$.
Equation~(\ref{eq:HHH}) is rewritten into
\begin{eqnarray}
	\sum_{m,\beta}\langle \gamma,k|H_F|\beta,m\rangle \Phi_{\alpha\beta}^{(m)}=\varepsilon_\alpha \Phi_{\alpha\gamma}^{(k)}\;,
\end{eqnarray}
with the entry of $H_F$ being
\begin{eqnarray}\label{eq:element}
	\langle\gamma,k|H_F|\beta,m\rangle=H_{\gamma \beta}^{(k-m)}+k\omega \delta_{km}  \delta_{\gamma \beta} \;,
\end{eqnarray}
where the term of $``k\omega"$ corresponds to the multi-photon process with $\hbar=1$.
The infinite-dimensional block matrix form of $H_F$ writes 
\begin{eqnarray}\label{eq:HF}
	H_F \!=\! \left(\begin{matrix}
		\ddots&\vdots&\vdots &\vdots&\begin{sideways} $\ddots$  \end{sideways}\\
		\cdots&\mathscr{H}^{(0)}-\omega I & \mathscr{H}^{(1)} & \mathscr{H}^{(2)}&\cdots\\
		\cdots&\mathscr{H}^{(-1)} & \mathscr{H}^{(0)} & \mathscr{H}^{(1)}&\cdots\\
		\cdots&\mathscr{H}^{(-2)} & \mathscr{H}^{(-1)} & \mathscr{H}^{(0)}+\omega I&\cdots\\
		\begin{sideways} $\ddots$  \end{sideways}&\vdots	&\vdots &\vdots&\ddots\\
	\end{matrix}\right),
\end{eqnarray}
where $I$ denotes the identity operator. 
This Floquet matrix is numerically truncated to a finite size $N(2n+1)$ in the Sambe space, which corresponds to an $N$-level quantum system with Fourier indices belonging to $\{0,\pm1,\cdots \pm n\}$.
A Floquet sector refers to the subspace of the Sambe space associated with a fixed Fourier index.
Different sectors are coupled by the nonzero Fourier components of the time-periodic Hamiltonian.
So we can see that the diagonal block matrices of equation~(\ref{eq:HF}) denote the static component dressed by integer multiples of the driving frequency in the same Floquet sector.
In contrast, the off-diagonal blocks describe the coupling between diﬀerent Floquet sectors, which can give rise to the distinct topological structures~\cite{Floquetinsulator}. 
We study a 2-level periodic system $H(t)=-\frac{\Delta}{2}\sigma_x-\frac{A \cos(\omega t)+\epsilon_0}{2}\sigma_z$ as the toy model.
By using the eigenbasis $\{|0\rangle,|1\rangle\}$ and the Fourier components $H^{(n)}$ of equation~(\ref{eq:Fourier}), we can obtain 
\begin{align*}
	\mathscr{H}^{(0)}
	&=\left(\begin{matrix}
		{H}_{0 0}^{(0-0)} &H_{0 1}^{(0-0)}\\
		H_{1 0}^{(0-0)} & H_{1 1}^{(0-0)}
	\end{matrix}\right)\nonumber\\
   &=
	\left(\begin{matrix}
		\langle0|H^{(0)}|0\rangle &\langle0|H^{(0)}|1\rangle\\
		\langle1|H^{(0)}|0\rangle & \langle0|H^{(0)}|1\rangle
	\end{matrix}\right)
    =-\frac{1}{2}\left(\begin{matrix}
		\epsilon_0& \Delta\\
		\Delta & -\epsilon_0
	\end{matrix}\right),\nonumber\\
	\mathscr{H}^{(1)}
	&=\left(\begin{matrix} 
		H_{0 0}^{(0-(-1))} & H_{0 1}^{(0-(-1))}\\
		H_{1 0}^{(0-(-1))} & H_{1 1}^{(0-(-1))}
	\end{matrix}\right) \nonumber\\
   &= \left(\begin{matrix}
		\langle 0|H^{(1)}|0\rangle & \langle 0|H^{(1)}|1\rangle\\
		\langle 1|H^{(1)}|0\rangle & \langle 1|H^{(1)}|1\rangle
	\end{matrix}\right)
=-\frac{1}{4}\left(\begin{matrix}
		A & 0\\
		0& -A
	\end{matrix}\right),\nonumber\\
	\mathscr{H}^{(-1)}
	&=\left(\begin{matrix} 
		H_{0 0}^{(0-1)} & H_{0 1}^{(0-1)}\\
		H_{1 0}^{(0-1)} & H_{1 1}^{(0-1)}
	\end{matrix}\right) \nonumber\\
&= \left(\begin{matrix}
		\langle 0|H^{(-1)}|0\rangle & \langle 0|H^{(-1)}|1\rangle\\
		\langle 1|H^{(-1)}|0\rangle & \langle 1|H^{(-1)}|1\rangle
	\end{matrix}\right)
    =-\frac{1}{4}\left(\begin{matrix}
		A & 0\\
		0& -A
	\end{matrix}\right),
\end{align*}
and other blocks are zero matrices.

In the Hilbert space $\bm H$, the Floquet mode satisfies $|\Phi_\alpha(t)\rangle=U(t)|\Phi_\alpha(0)\rangle$ where the time evolution operator obeys the Schr\"odinger  equation $i\partial U(t)/\partial t=H(t) U(t)$. Therefore, the time evolution operator writes
$
U(t)=\mathcal{T} e^{-i \int_0^t ds H(s)},
$
where $\mathcal{T}$ is the time-order operator. 
By expanding the Hilbert space $\bm{H}$ to the Sambe space $\bm{S}$, the time-periodic Hamiltonian $H(t)$ is mapped into the time-independent Floquet Hamiltonian $H_F$, the time-evolution operator becomes
\begin{eqnarray}\label{eq:Time}
	U_F(t)=e^{-iH_Ft},
\end{eqnarray}
where the evolution time $t$ is in a stroboscopic fashion in steps of the driving period $T$,  which implies the micromotion operator being an identity operator~\cite{micromotion,FloquetReview}.
The exact diagonalization (ED) for the truncated $H_F$ is recorded as
\begin{eqnarray}\label{eq:eigen}
	H_F=D \Lambda D^{\dagger}=\sum_{\alpha=1}^{N(2n+1)} \lambda_\alpha |\lambda_\alpha\rangle \rangle \langle \langle \lambda_\alpha|\;,
\end{eqnarray} 
with the matrices
\begin{eqnarray}
	\Lambda&=&\text{diag}\left(\lambda_1,\lambda_2,\cdots \lambda_{N(2n+1)}\right),\nonumber\\
	D&=&\left( |\lambda_1\rangle\rangle,|\lambda_2\rangle\rangle,\cdots,|\lambda_{N(2n+1)}\rangle\rangle\right)\;,
\end{eqnarray}
where the diagonal matrix $\Lambda$ contains all the eigenvalues $\{\lambda_\alpha\}$. 
The columns of the matrix $D$ are the Floquet eigenstates in the Sambe space and satisfy $|\lambda_\alpha\rangle \rangle=\sum_{\beta,m}|\beta,m\rangle \langle \beta,m|\lambda_\alpha\rangle\rangle$ for $\sum_\alpha |\lambda_\alpha\rangle \rangle \langle \langle \lambda_\alpha|=I$.
By inserting equations~(\ref{eq:eigen}) into (\ref{eq:Time}), we get the entry of the time evolution operator as 
\begin{eqnarray}\label{eq:UU}
	{\hspace{-6mm}}\langle \gamma,k|U_F(t)|\beta,m\rangle = \sum_\alpha \langle \gamma,k| \lambda_\alpha\rangle\rangle \langle\langle \lambda_\alpha   |\beta,m\rangle e^{-i\lambda_\alpha t}.
\end{eqnarray}
Equation~(\ref{eq:UU}) can be mapped back onto the Hilbert space as 
\begin{eqnarray}\label{eq:Ut}
	\langle \gamma|U(t)|\beta\rangle&=&\sum_k \langle \gamma,k|U_F(t)|\beta,0\rangle e^{ik\omega t} \nonumber\\
	&=&\sum_{\alpha,k} B_{\alpha k}  e^{-i\lambda_\alpha t} e^{ik\omega t}\;, 
\end{eqnarray}
with 
\begin{eqnarray}
	B_{\alpha k}=\langle \gamma,k|\lambda_\alpha\rangle \rangle \langle \langle \lambda_\alpha|\beta,0\rangle.
\end{eqnarray}
Equation~(\ref{eq:Ut}) can be understood as the transition amplitude that the system initially in the state $|\beta,0\rangle$～\cite{Explain}
and evolves to the  $|\gamma,k\rangle$ at time $t$, which is summed over $k$ with the phase factor $e^{ik\omega t}$.
Accordingly, the corresponding transition probability with different Fourier indices $k$ and $k'$ writes
\begin{eqnarray}\label{eq:P}
	P_{\beta\gamma}(t) &=& |\langle \gamma|U(t)|\beta\rangle|^2 \nonumber\\
	&=&\sum_k|C_k(t)|^2 \!+\! \sum_{k\neq k'} C_k(t) C_{k'}^*(t) e^{i(k-k')\omega t},
\end{eqnarray}
where $C_{k(k')}(t)=\langle \gamma,k(k')| U_F(t) |\beta,0\rangle=\sum_{\alpha,k(k')} B_{\alpha k(k')} e^{-i\lambda_\alpha t}$  is defined, respectively.
The first term of equation~(\ref{eq:P}) denotes the probability summation contributed by each Floquet sideband from the same sector, and the second term comes from the interference among different Floquet sidebands from different sectors with the phase factor $e^{i(k-k')\omega t}$.

Two distinct forms of averaged transition probabilities are commonly defined: (1) Following Shirley’s formulation~\cite{Shirley1965}, a period-averaged transition probability eliminates the hybridization between different Floquet sidebands but the coherence between quasienergy spectrums in the same sector survives, which yields ${P}_{\beta\gamma}^{(1)}(t)=\sum_{k} |C_k(t)|^2=\sum_{\alpha,\alpha',k}B_{\alpha k} B_{\alpha' k}^* e^{-i(\lambda_\alpha-\lambda_{\alpha'})t}$ 
since $\frac{1}{T}\int_t^{t+T} ds e^{i(k-k')\omega s}=\delta_{kk'}$. 
The same averaged transition probability is also presented in Ref.~\cite{An2023}.
(2) In the long-time limitation $T \to \infty$, $(\lambda_\alpha-\lambda_{\alpha'})-(k-k')\omega=0$ is satisfied and only resonance terms survive, which results in ${P}_{\beta\gamma}^{(2)}(t)=\sum_{\alpha,k} |B_{\alpha k}|^2$.
The same averaged transition probability is also presented in Ref.~\cite{FF}.

\subsection{Quantum multiparameter estimation of effective Floquet Hamiltonian}\label{Subsec:Fparameter}
For multiple to-be-estimated parameters ${\bf x}=\{x_1,x_2,\cdots\}$ encoded in the time-periodic Hamiltonian $H(t)$, the estimation precision is quantified by the error covariant matrix $\bm{Cov}( \tilde{\mathbf{x}})$ of the local unbiased estimator $\tilde{\mathbf{x}}=\{\tilde{x}_1,\tilde{x}_2,\cdots\}$. 
The matrix element writes $\left[\bm{Cov}(\tilde{\mathbf{x}})\right]_{\ell \ell'}=\text{E}[\tilde{x}_\ell \tilde{x}_{\ell'}]-\text{E}[\tilde{x}_\ell] \text{E}[\tilde{x}_{\ell'}]$ ($\text{E}[\bullet]$ denotes the expectation).
The highest estimation precision is described by the quantum Cram$\acute{e}$r-Rao bound (QCRB)~\cite{Paris2009,Control,Multi2020}.  
The QCRB inequality shows $\bm{Cov}({\bf x}) \ge \nu^{-1} \mathcal{\mathbf{F}}^{-1} \ge \nu^{-1} \mathcal{\mathbf{I}}^{-1}$, where $\nu$ denotes the repetition number of the measurement, the matrix $\mathbf{F}$ represents the classical Fisher information matrix (CFIM), and the matrix $\mathbf{I}$ indicates the quantum Fisher information matrix (QFIM). 
The $\ell$-th diagonal element of the QFIM or the CFIM corresponds to the Quantum Fisher information (QFI) $I_\ell$ or the Classical Fisher information (CFI) $F_\ell$ for the unknown parameter $x_\ell$.

Within the framework of Floquet theory and quantum multiparameter estimation, the CFI takes the form
\begin{eqnarray}\label{eq:CFI}
	\mathcal{F}_\ell=\sum_{\gamma} \frac{1}{P_{\beta \gamma}(t)} \left( \frac{\partial P_{\beta \gamma}(t)} {\partial x_\ell} \right)^2\;,
\end{eqnarray}
where the transition probability $P_{\beta \gamma}(t)$ is deduced from equation~(\ref{eq:P}) and is accessible through stroboscopic projective measurements~\cite{M1,M2}.
By inserting equations~(\ref{eq:Ut})-(\ref{eq:P}) into (\ref{eq:CFI}), the CFI can be further decomposed as
\begin{widetext}
\begin{eqnarray}\label{eq:Fell}
	\mathcal{F}_\ell = \sum_\gamma \! \frac{\! 4\text{Re}^2 \! \Big[D_{\beta\gamma}^*(t)  \sum_{\alpha,k} \Big(
		\overbrace{ \frac{\partial B_{\alpha k}}{\partial x_\ell}}^{\substack{\text{Floquet} \\ \text{eigenmodes}}}
		\overbrace{- i t B_{\alpha k} \frac{\partial \lambda_\alpha}{\partial x_\ell}}^{\text{Quasienergies}} 
		\overbrace{+ikt B_{\alpha k} \frac{\partial \omega}{\partial x_\ell}}^{\substack{\text{Multi-photon} \\ \text{process}}}   
		\Big)  e^{-i\lambda_\alpha t} e^{ik\omega t}  \Big]}{|D_{\beta \gamma} ({t})|^2}.\nonumber\\
\end{eqnarray}
\end{widetext}
where $D_{\beta\gamma}(t)=\langle \gamma|U(t)|\beta\rangle$ and $\text{Re}[\bullet]$ represents the real part.
The hints
$P_{\beta \gamma}(t)=|D_{\beta\gamma}(t)|^2$
and ${\partial P_{\beta \gamma}(t)}/{\partial x_\ell}=2\text{Re}\left[D_{\beta\gamma}^*(t) {\partial D_{\beta\gamma}(t)}/{\partial x_\ell}\right]$ are employed.
Equation~(\ref{eq:Fell}) straightforwardly indicates that the influence of infinitesimal variations of the parameter $x_\ell$ on measurement precision stems from three distinct contributions: one arising from the Floquet eigenmodes via $\partial B_{\alpha k}/\partial x_\ell$, the other from the quasienergies via $\partial \lambda_\alpha/\partial x_\ell$, and the rest from the multi-photon process via $\partial \omega/\partial x_\ell$.

The generator is responsible for the unitary evolution induced by the infinitesimal variations of the parameter $x_\ell$, which is defined as
\begin{eqnarray}\label{eq:gen}
	h_\ell(t)=i \left( \frac{\partial U(t)}{\partial x_\ell} \right) U^\dagger(t)\;.
\end{eqnarray}
The matrix element of the generator is
$\langle \gamma|h_\ell(t)|\beta\rangle 
= i \sum_\mu \left({\partial \langle \gamma|U(t)|\mu\rangle}/{\partial x_\ell}\right) \langle \mu|U^\dagger(t)|\beta\rangle$, which further simplifies to $ i \sum_\mu D_{\mu \beta}^*(t) {\partial D_{\mu \gamma}(t)}/{\partial x_\ell}$.
The generator naturally comprises three distinctive components:
\begin{eqnarray}
	h_\ell(t)=\mathfrak{h}_{\ell,\alpha}(t)+\mathfrak{h}_{\ell,\varepsilon}(t)+\mathfrak{h}_{\ell,\omega}(t).
\end{eqnarray}
Their matrix elements contributed by Floquet eigenmodes, quasienergies, and multi-photon processes, respectively, are
\begin{eqnarray}\label{eq:ele}
	&&{\hspace{-0.5cm}}\langle \gamma|\mathfrak{h}_{\ell,\alpha}(t)|\beta\rangle= i \sum_{\mu,\alpha,k} \frac{\partial \widetilde{B}_{\alpha k}}{\partial x_\ell} e^{-i\lambda_\alpha t} e^{ik\omega t}  D_{\mu \beta}^*(t),
	\nonumber\\
	&&{\hspace{-0.5cm}}\langle \gamma|\mathfrak{h}_{\ell,\varepsilon}(t)|\beta\rangle=t \sum_{\mu,\alpha,k} \widetilde{B}_{\alpha k}\frac{\partial \lambda_\alpha}{\partial x_\ell} e^{-i\lambda_\alpha t} e^{ik\omega t} D_{\mu \beta}^*(t),  \nonumber\\
	&&{\hspace{-0.5cm}}\langle \gamma|\mathfrak{h}_{\ell,\omega}(t)|\beta\rangle \!=\! -kt \sum_{\mu,\alpha,k} \widetilde{B}_{\alpha k} \frac{\partial \omega}{\partial x_\ell} e^{-i\lambda_\alpha t} e^{ik\omega t}D_{\mu \beta}^*(t),
\end{eqnarray}
where $\widetilde{B}_{\alpha k}=\langle \gamma,k|\lambda_\alpha\rangle \rangle \langle \langle \lambda_\alpha|\mu,0\rangle$.
The fluctuations of the generator associated with the initial probe state $\rho_0=|\Psi(0)\rangle\langle\Psi(0)|$ yield the QFI~\cite{Paris2009,Control,Multi2020,Pang2017,Hou2021}.  
Accordingly, the QFI is constituted by
\begin{widetext}
\begin{align}\label{eq:Ig}
	\mathcal{I}_{\ell}&=4( \text{Tr}\left[{h}_{\ell}^2(t) \rho_0\right]-\text{Tr}^2\left[{h}_{\ell}(t) \rho_0\right])\nonumber\\
	&=\overbrace{{I}_{\ell,\alpha}}^{\substack{\text{Floquet} \\ \text{eigenmodes}}}
	\overbrace{+{I}_{\ell,\varepsilon}}^{\text{Quasienergies}} 
	\overbrace{+{I}_{\ell,\omega}}^{\substack{\text{Multi-photon} \\ \text{process}}}  
	\overbrace{+8(\text{Cov}[\mathfrak{h}_{\ell,\alpha},\mathfrak{h}_{\ell,\varepsilon}]+\text{Cov}[\mathfrak{h}_{\ell,\alpha},\mathfrak{h}_{\ell,\omega}]+\text{Cov}[\mathfrak{h}_{\ell,\varepsilon},\mathfrak{h}_{\ell,\omega}])}^{\text{Coherence}},
\end{align}
\end{widetext}
with the components obtained from Floquet eigenmodes, quasienergies, and multi-photon processes as
\begin{eqnarray}\label{eq:QFIcom}
	&&{I}_{\ell,\alpha}=4( \text{Tr}\left[\mathfrak{h}_{\ell,\alpha}^2(t) \rho_0\right]-\text{Tr}^2\left[\mathfrak{h}_{\ell,\alpha}(t) \rho_0\right] )\;,\nonumber\\
	&&{I}_{\ell,\varepsilon}=4( \text{Tr}\left[\mathfrak{h}_{\ell,\varepsilon}^2(t) \rho_0\right]-\text{Tr}^2\left[\mathfrak{h}_{\ell,\varepsilon}(t) \rho_0\right] )\;,\nonumber\\
	&&{I}_{\ell,\omega}=4 ( \text{Tr}\left[\mathfrak{h}_{\ell,\omega}^2(t) \rho_0\right]-\text{Tr}^2\left[\mathfrak{h}_{\ell,\omega}(t) \rho_0\right] ),
\end{eqnarray}
where $\text{Tr}[\bullet]$ and $\text{Cov}[\bullet,\bullet]$ individually represent the matrix trace and the covariance of matrices.
By investigating the maximal spread of the generator, we further arrive at the upper bound of the QFI
\begin{eqnarray}\label{eq:IMax}
	\mathcal{I}_\ell \le \mathcal{I}_\ell^{M}\!=\!\left\{\text{Max}[h_\ell(t)]-\text{Min}[h_\ell(t)]\right\}^2,
\end{eqnarray}
where $\text{Max/Min}[\bullet]$ denote the maximal/minimal eigenvalue of the operator, respectively.

For quantum single-parameter estimation, the QCRB inequality can be approximately saturated by employing the maximum likelihood estimation (MLE)~\cite{Paris2009}, while for the multiparameter case it is not always attainable~\cite{Control,Multi2020,PRX2021}. This unattainability comes from the fact that generators for the different parameters are non-commutative in general, which induces the trade-off among estimation precisions. 
This limitation is known as {\it measurement incompatibility}~\cite{IM1,IM2,IM3,IM4}. Its absence for different parameters $x_\ell,x_{\ell'}$ can be demonstrated by the weak commutation condition
\begin{eqnarray}\label{eq:Mu}
	\Omega_{\ell \ell'}=\text{Im}[\text{Tr}\left[\rho_0 [h_\ell(t),h_{\ell'}(t)]\right]]=0\;,
\end{eqnarray}
where $[\bullet,\bullet]$ denotes the commutativity of operators, and $\text{Im}[\bullet]$ extracts the imaginary part.
The primary objective of quantum multiparameter estimation is to fulfill the simultaneous optimal estimation for different parameters.
This requires that equations~(\ref{eq:IMax})-(\ref{eq:Mu}) are satisfied concurrently with respect to all the to-be-estimated parameters, which can be guaranteed at the TPT as presented in the following section.\\

\section{Ring-shaped Rashba spin-orbit interferometer model}\label{Sec:Rashba}
In this section, we consider a ring-shaped Rashba spin-orbit interferometer~\cite{Lyanda1993,Rashba2015,NJP2017,PRR2020} to demonstrate the proposed Floquet parameter estimation framework. 
This model features a periodically driven-induced topological phase transition, whose criticality provides a platform for the Floquet-enhanced quantum sensing.
\begin{figure}[!h]
	\centering
	\includegraphics[width=0.45\textwidth]{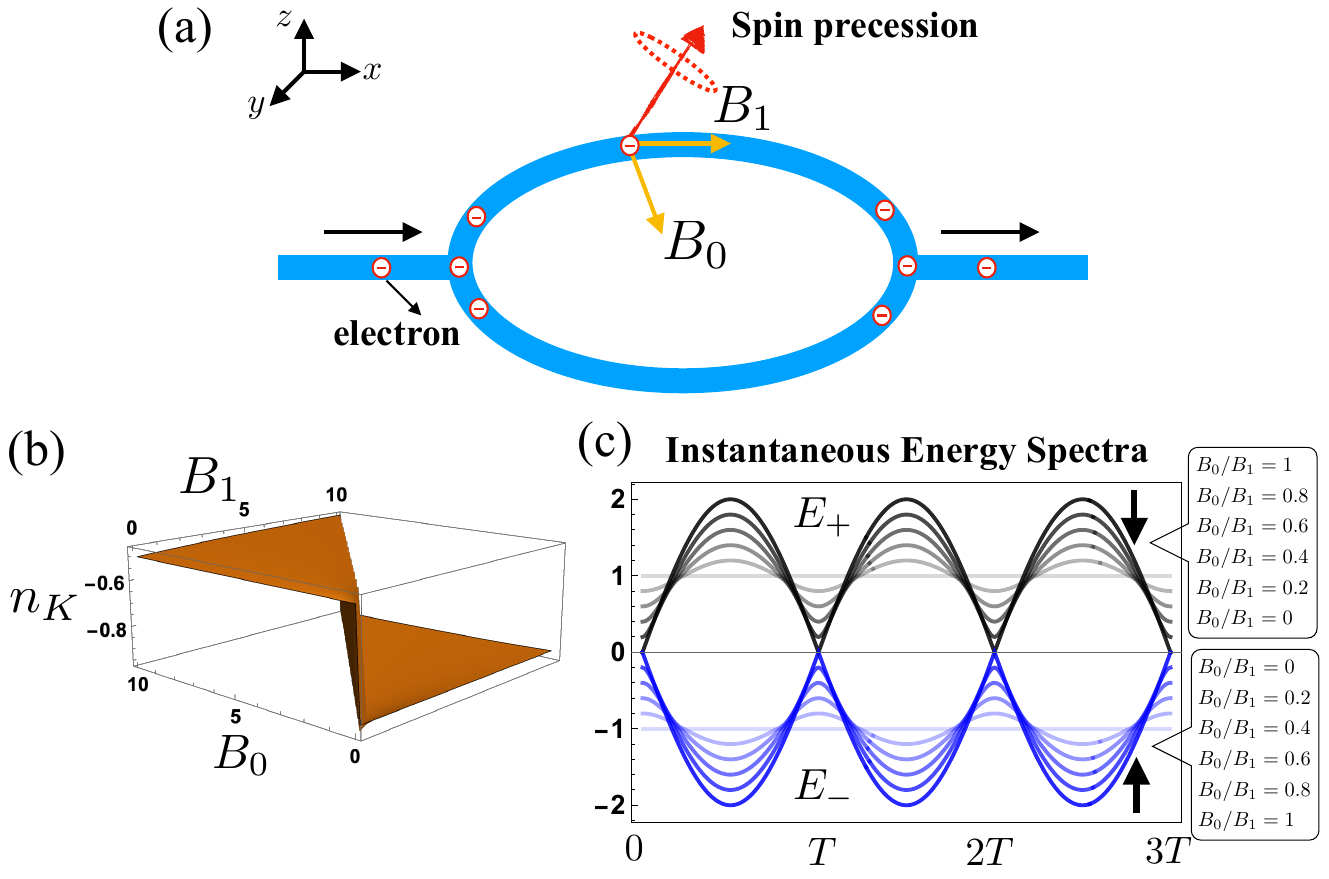}
	\caption
	{(a) Ring-shaped Rashba spin-orbit interferometer.
		The system is subjected to an $xy$-plane Rashba magnetic field $B_0$ and a Zeeman field $B_1$ along the $x$-direction. The motion of electrons accompanies the spin precession. 
		(b) Winding number $n_K$ of equation~(\ref{eq:newnK}) versus the parameters $B_0$ and $B_1$.
		(c) Instantaneous energy spectra of the Hamiltonian (\ref{eq:H}) during three time periods.
		The top and bottom halves respectively show the behavior of the instantaneous eigenvalues $E_\pm=\pm \sqrt{B_0^2+B_1^2-2 B_0 B_1 \cos(\omega t)}$ near the topological phase transition (TPT). The different spectrum curves correspond to distinct ratios between $B_0$ and $B_1$.}
	\label{Figscheme}
\end{figure}

As shown in figure~\ref{Figscheme} (a),  a mesoscopic ring conductor is composed by a two-dimensional electron gas (2DEG)  in a layered semiconductor quantum well (QW) such as InP/InGaAs/InAlAs.
A top-gate electrode applies the alternating voltages to the QW, which breaks the spatial inversion symmetry of the QW so as to yield a macroscopic electric field. 
The confined electrons in the 2DEG  move across this electric field along the ring, which yields an effective Rashba magnetic field with the strength $B_0$ and induces the spin precession.  
The precession angle is associated with the strength of the Rashba spin-orbit interaction (SOI) and is embodied in the geometrical phase.
The Zeeman magnetic field with the strength $B_1$ acts on the 2DEG as well.
The Hamiltonian of this ring-shaped Rashba spin-orbit interferometer is~\cite{NJP2017}
\begin{eqnarray}\label{eq:H}
	H(t)=B_0 \left[\cos(\omega t)\sigma_x+ \sin(\omega t)\sigma_y\right] -B_1\sigma_x,
\end{eqnarray}
where the Rashba magnetic field is in the $xy$-plane, the Zeeman magnetic field is along the $x$-direction, and $\{\sigma_x,\sigma_y,\sigma_z\}$ denote the Pauli operators. 
The SOI dominates the evolution of the spin state for a rotating field when $B_0/B_1>1$  and encloses the point of vanishing field in its round trip. 
For the rotating field, the rotation wave approximation (RWA) can be used to studying the transition probability of a particle from the  ground state to the excited state. 
The quantum dynamics evolution follows the adiabatic approximation.
In contrast,  the Zemman magnetic field dominates the evolution of the spin state for an oscillating field when $B_0/B_1<1$ and leaves the point of vanishing field out of the loop. 
For the oscillating field, the RWA is invalid since that the counter-rotating term will cause the Bloch-Siegert shift of the resonance frequency~\cite{Nagasawa2013}. In this case, the dynamics evolution is non-adiabatic~\cite{PRR2020,Lyanda1993,NJP2017,Nagasawa2013}.
Especially at $B_0/B_1=1$, a topological phase transition emerges in this hybrid Rashba-Zeeman field.

\subsection{Winding number}\label{Subsec:winding}
To identify the TPT, equation~(\ref{eq:H}) can be rewritten into
\begin{eqnarray}\label{eq:genH}
	H(t)=\mathcal{B}(t)[\cos{\vartheta}(t) \sigma_x+\sin\vartheta(t)\sigma_y]\;, 
\end{eqnarray}
with
\begin{align}
	\mathcal{B}(t)&=\sqrt{B_0^2+B_1^2-2B_0 B_1 \cos(\omega t)}\;,\label{eq:Bt1}\\
	\vartheta(t)&=\arctan\left[\frac{\sin(\omega t)}{\cos(\omega t)-\frac{B_1}{B_0}}\right]\;.\label{eq:Bt}
\end{align}
The driving curvature of the applied field is defined as~\cite{PRR2020}
\begin{eqnarray}\label{eq:curvature}
	K(t)=-\partial_t \vartheta(t)\;,
\end{eqnarray}
where $\partial_t(\bullet)=\partial(\bullet)/\partial t$.
The varied driving curvature $K(t)$ can be achieved by deforming the geometry of a one-dimensional semiconducting nanostructure~\cite{Ying2016}.
The integral of the curvature over one time period $T$  gives the winding number 
\begin{eqnarray}\label{eq:nk}
	n_K=\frac{1}{2\pi} \int_0^T dt K(t)\;.
\end{eqnarray}
By substituting equations~(\ref{eq:Bt})-(\ref{eq:curvature}) into (\ref{eq:nk}), we get
\begin{eqnarray}\label{eq:newnK}
	n_K&=&\frac{-T \omega}{4 \pi}+\frac{1}{2\pi}\arctan\left[\frac{(B_0-B_1)\cot\left(\frac{T \omega}{2}\right)}{B_0+B_1}\right] \nonumber\\
	&-&\frac{1}{4 \text{sign}(B_0-B_1)}\;.
\end{eqnarray}
We plot the winding number of equation~(\ref{eq:newnK}) in figure~\ref{Figscheme} (b).
The result of $n_K=\left\{\begin{matrix}
	0, B_1 <B_0\\
	-1, B_1> B_0\end{matrix}\right.$ reveals the TPT occurring at $B_0=B_1$.
Moreover, by analyzing the instantaneous energy spectra displayed in figure~\ref{Figscheme} (c), we can see that the energy gap closes if and only if $B_0/B_1=1$ at the integer multiples of the period.
The crossing of the degeneracy point breaks down the adiabatic approximation~\cite{Aharonov1987}.
The spin texture features of this model are also shortly discussed in appendix~A. Besides, one can also numerically investigate the quasienergy spectrum to identify the TPT.
Exact diagonalization for the Floquet matrix of equation~(\ref{eq:HF}) gives a set of eigenvalues $\{\lambda_\alpha\}$ that can be expressed as $\{\varepsilon_\alpha+ n\omega \}$ and correspond to the Floquet sidebands in the different sectors.
Since quasienergies differing by integer multiples of $\omega$ are physically equivalent, one usually considers those within a given zone. For example, the first Brillouin zone (FBZ) restricts $\tilde{\varepsilon}_\alpha=(\varepsilon_\alpha+n\omega) (\text{mod}\; \omega) \in (-\frac{\omega}{2},\frac{\omega}{2}]$~\cite{Flo1,PRXFloquet}.
The set $\{\tilde{\varepsilon}_\alpha\}$ then forms the quasienergy spectra. The location of gap closing in the quasienergy spectrum corresponds to the presence of the TPT.

\subsection{Total phase signature}\label{Subsec:phase}
For the presented ring-shaped model,
we also respectively analyze the Berry phase in the adiabatic approximation and the Aharonov-Anandan phase in the non-adiabatic evolution.
For $B_0/B_1>1$, the point of vanishing field is enclosed, and the spin evolution follows the adiabatic approximation.
Whereas for $B_0/B_1<1$, the loop excludes this point, and the evolution becomes nonadiabatic. 
The TPT at $B_0/B_1=1$ can be identified from the nonadiabatic evolution in the rotating frame.
The sum of geometric and dynamical phases (i.e., the total phase) over a periodic evolution is a key signature, the resulting manifestation of the TPT is plotted in figure~\ref{Fig_totalphase}.

\subsubsection{Berry phase in adiabatic approximation}
For a generic spin state $|\psi(t)\rangle$, the Berry phase is defined as
\begin{eqnarray}\label{eq:Berryphase}
	\gamma_B=\int_0^Tdt  \langle\psi(t)|i  {\partial_t |\psi(t)\rangle}=\frac{\Omega}{2}\;,
\end{eqnarray} 
with the solid angle
\begin{eqnarray}\label{eq:solidangle}
	\Omega=2\pi(1-\cos\theta)\;,
\end{eqnarray}
where $\theta$ represents the angle between the spin precession axis and the binormal direction to the ring plane.
The ratio $\tan \theta=\omega_\text{SO}/\omega$ relates the Lamor frequency $\omega_\text{SO}$ of spin precession to the frequency $\omega$ of changing the magnetic field direction.
When the spin precession is fast enough compared with the propagation of the electrons along the ring (i.e., $\omega_\text{SO} \gg \omega$ in the adiabatic limit~\cite{Richter2003}), the spin precession axis is parallel to the direction of the magnetic field induced by the SOI so that $\theta=\pi/2$~\cite{Nagasawa2012}.  
Based on equations~(\ref{eq:Berryphase})-(\ref{eq:solidangle}), we get
\begin{eqnarray}\label{eq:BP}
	\gamma_B=\pi\;.
\end{eqnarray}

\subsubsection{Aharonov-Anandan phase in non-adiabatic evolution}
\begin{figure}[!h]
	\centering
	\includegraphics[width=0.35\textwidth]{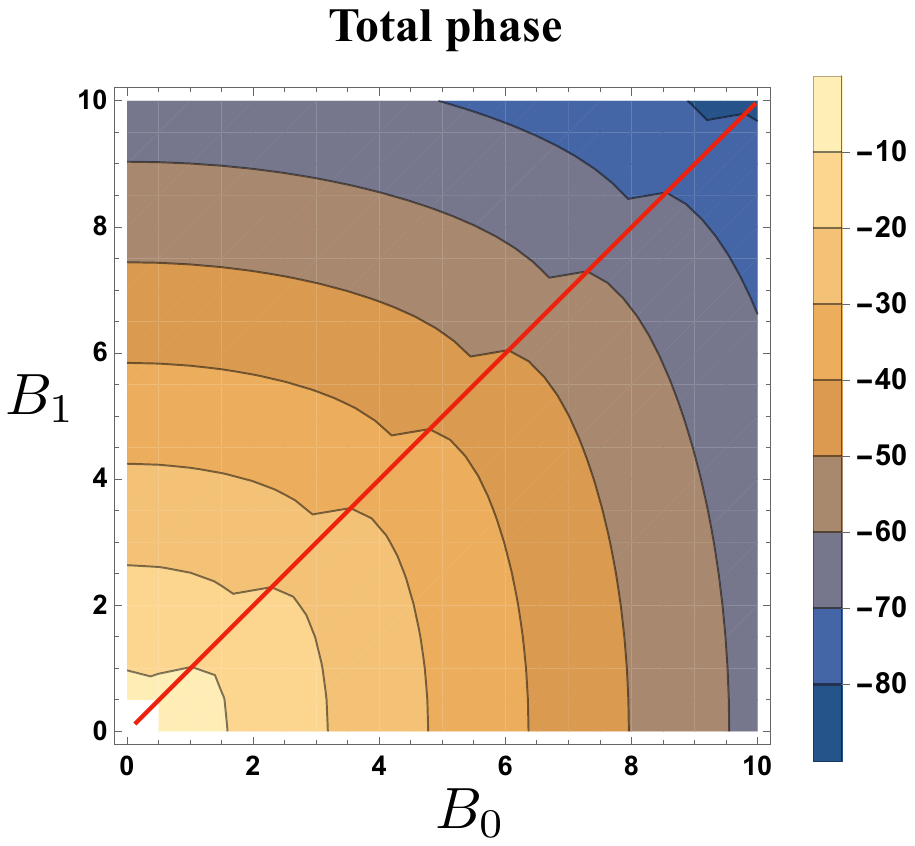}
	\caption
	{Contourplot of the total phase of equation~(\ref{eq:tot}). The red line indicates the boundary of the topology phase transition.}
	\label{Fig_totalphase}
\end{figure}

One can perform  an SU(2) rotation transformation such that the spin is instantaneously aligned with the applied field.  
The generic spin state $|\psi(t)\rangle$ is transformed into 
$|\psi_R(t)\rangle=U^\dagger (t) |\psi(t)\rangle$ with $U^\dagger(t)=\text{exp}\left(i \vartheta(t) \sigma_z/2 \right)$.
The Schr\"odinger equation for $|\psi_R(t)\rangle$ writes 
\begin{eqnarray}
	i\hbar {\partial_t |\psi_R(t)\rangle}&=&\left[U^\dagger(t) H(t) U(t)+i {\partial_t U^\dagger(t)} U(t)\right]  |\psi_R(t)\rangle \nonumber\\
	&=&\left[\mathcal{B}(t) \sigma_x+\frac{K(t)}{2} \sigma_z\right]|\psi_R(t)\rangle \nonumber\\
	&=&\widetilde{H}(t)|\psi_R(t)\rangle\;,
\end{eqnarray}
where equations~(\ref{eq:Bt1})-(\ref{eq:curvature}) are utilized.
If the rotated spin state $|\psi_R(t)\rangle$ is exactly the eigenstate $|\psi_{RA}(t)\rangle$ of $\widetilde{H}(t)$, the nonadiabatic spin state is $|\widetilde{\psi}(t)\rangle=U(1)U(t)|\psi_{RA}(t)\rangle=\widetilde{U}(t)|\psi_{RA}(t)\rangle$ with $U(1)=\text{exp}\left(i \vartheta(t) /2\right)$~\cite{PRR2020,Aharonov1987}.
The Aharonov-Anandan geometric phase, the non-adiabatic analogue of the Berry phase, writes
\begin{eqnarray}\label{eq:gephase}
	\gamma_A&=&\int_0^T dt \langle\widetilde{\psi}(t)|i \partial_t |\widetilde{\psi}(t)\rangle \nonumber\\
	&=&\int_0^Tdt \langle\psi_{RA}(t)|\widetilde{U}^\dagger(t) i \partial_t \widetilde{U}(t)|\psi_{RA}(t)\rangle + \widetilde{\gamma}_B,
\end{eqnarray}
with
\begin{eqnarray}\label{eq:BeP}
	\widetilde{\gamma}_B=\int_0^Tdt  \langle\psi_{RA}(t)|i\partial_t|\psi_{RA}(t)\rangle\;.
\end{eqnarray} 
The eigenstate of $\widetilde{H}(t)$ takes the form of $|\psi_{RA}(t)\rangle=\left(\begin{matrix}
	\cos(\alpha/2)\\ e^{i\phi} \sin(\alpha/2)
\end{matrix}\right)$ with $\cos \alpha={{\hbar K(t)}}/{\sqrt{4\mathcal{B}^2(t)+\left({\hbar K(t)}\right)^2}}$ and $ e^{i\phi}=1$. 
This eigenstate form and equation~(\ref{eq:BeP}) lead to
\begin{eqnarray}
	\widetilde{\gamma}_B=0\;.\label{eq:num1}
\end{eqnarray}
The nonadiabatic dynamical phase is
\begin{eqnarray}\label{eq:dyphase}
	d&=&-\frac{1}{\hbar}\int_0^T dt \langle \widetilde{\psi}(t)|{H}(t)|\widetilde{\psi} (t)\rangle 	\nonumber\\
	&=&-\frac{1}{\hbar}\int_0^T dt \langle {\psi}_{RA}(t)| \widetilde{U}^\dagger(t){H}(t)\widetilde{U}(t)|{\psi}_{RA} (t)\rangle\nonumber\\
	&=&-\frac{\mathcal{B}(t)}{\hbar}\int_0^T dt \langle {\psi}_{RA}(t)| \sigma_x  |{\psi}_{RA} (t)\rangle\;.
\end{eqnarray}
The fact that $B_0/B_1<1$ makes the spin precession frequency $\omega_\text{SO}$ comparable to $\omega$ (non-adiabatic evolution).  The electron spin does not exactly precess around the magnetic field induced by the SOI, and there is an effective magnetic field normal to the ring's plane besides the SOI field so that $0<\theta<\pi/2$~\cite{Meijer2002,Nagasawa2013}.  
The in-plane field perturbs the eigenstate spin texture by reducing the solid angle.
The expressions for the geometric phase (\ref{eq:gephase}) and the dynamic phase (\ref{eq:dyphase}) are evaluated further by employing equations~(\ref{eq:Bt1})--(\ref{eq:curvature}), i.e.,
\begin{align}
	\gamma_A&=\frac{1}{2} \! \int_0^T dt K(t)\!-\!\frac{1}{2} \! \int_0^T \! dt \frac{\hbar K(t)^2}{\sqrt{4\mathcal{B}(t)^2+\hbar^2 K(t)^2}},\\
	d&=\frac{-1}{\hbar}\int_0^T dt \frac{2 \mathcal{B}(t)^2}{\sqrt{4 \mathcal{B}(t)^2+\hbar^2 K(t)^2}}.\label{eq:num3}
\end{align}
The total phase is
\begin{eqnarray}\label{eq:tot}
	\phi_\text{tot}=\gamma_A+d\;.
\end{eqnarray}
The numerical results of the total phase  (\ref{eq:tot}) are plotted in figure~\ref{Fig_totalphase}. It exhibits the total phase dislocation at $B_0=B_1$, which means the occurrence of the TPT.

\section{Floquet multiparameter estimation enhanced by topological phase transition}\label{Sec.FL}
For the ring-shaped Rashba spin-orbit interferometer of (\ref{eq:H}), the to-be-estimated parameters can be recognized with $\{B_0,B_1,\omega\}$.
When one simultaneously considers the estimation for these three parameters, it leads to a multiparameter estimation problem.
Our following numerical results reveal the pronounced peaks of the QFI for both $B_0$ and $B_1$ at the TPT with the HL $\sim t^2$. 
For the parameter $\omega$, the QFI scaling even reaches $\sim t^4$ at the TPT, which is beyond the HL.
This demonstrates the effectiveness of Floquet critical quantum sensing.

Based on equations~(\ref{eq:element})-(\ref{eq:HF}), we can obtain the Floquet Hamiltonian in terms of (\ref{eq:H}) as
\begin{widetext}
\begin{eqnarray}\label{eq:floquet}
	H_F=\left(\begin{array}{cccccccccccc}
		\ddots&\vdots&\vdots&\vdots&\vdots&\vdots&\vdots&\vdots&\vdots&\vdots&\vdots&\begin{sideways}
			$\ddots$
		\end{sideways}\\
		\cdots&-2\omega & -B_1& 0&B_0&0&0&0&0&0&0&\cdots\\
		\cdots&-B_1&-2\omega&0&0&0&0&0&0&0&0&\cdots\\
		\cdots&0&0&-\omega&-B_1&0&B_0&0&0&0&0&\cdots\\
		\cdots&B_0&0&-B_1&-\omega&0&0&0&0&0&0&\cdots\\
		\cdots&0&0&0&0&0 &-B_1&0&B_0&0&0&\cdots\\
		\cdots&0&0&B_0&0&-B_1&0&0&0&0&0&\cdots\\
		\cdots&0&0&0&0&0&0&\omega&-B_1&0&B_0&\cdots\\
		\cdots&0&0&0&0&B_0&0&-B_1&\omega&0&0&\cdots\\
		\cdots&0&0&0&0&0&0&0&0&2\omega&-B_1&\cdots\\
		\cdots&0&0&0&0&0&0&B_0&0&-B_1&2\omega&\cdots\\
		\begin{sideways}
			$\ddots$
		\end{sideways}&\vdots&\vdots&\vdots&\vdots&\vdots&\vdots&\vdots&\vdots&\vdots&\vdots&\ddots\\
	\end{array}\right)\;,
\end{eqnarray}
\end{widetext}
where for brevity we omit the other matrix elements and display only those for $k,m=\{0,\pm 1, \pm 2\}$ (see the indices in equation~(\ref{eq:element})). 
By diagonalizing the Floquet Hamiltonian of (\ref{eq:floquet}), the time evolution operator of equation~(\ref{eq:Ut}) can be determinated. 
Sequentially, the generator of equation~(\ref{eq:gen}), the QFI of equation~(\ref{eq:Ig}), the QFI upper bound of equation~(\ref{eq:IMax}), and the measurement incompatibility recognized with equation~(\ref{eq:Mu}) are finally obtained by using the center difference method (CDM).
Especially at the stroboscopic measurement time $t=lT_0=2l\pi/\omega_0$ for $l \in \mathbb{N}+$ with $\omega_0$ being the external clock frequency, both the generator and the resulting QFI components reduce to simpler forms (see appendix~B).

\subsection{Component-resolved QFI and scaling behavior}
\begin{figure*}[htp]
	\centering
	\includegraphics[width=0.85\textwidth]{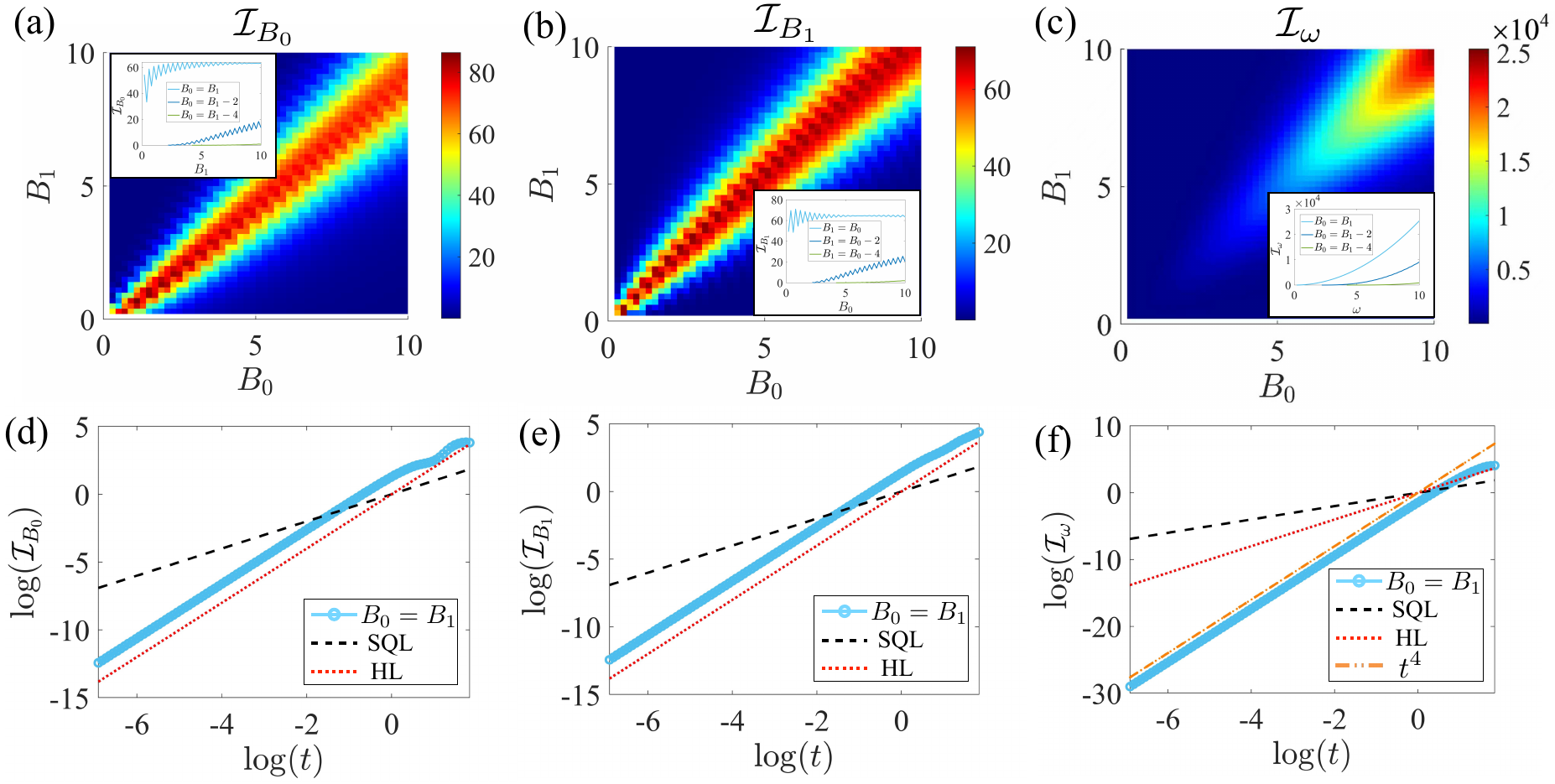}
	\caption
	{(a)-(c) QFIs $\mathcal{I}_{B_0}$, $\mathcal{I}_{B_1}$, $\mathcal{I}_\omega$ for the parameters $B_0$, $B_1$, and $\omega$ versus $B_0$ and $B_1$.
		The insets in panels (a)-(c) illustrate the behavior of $\mathcal{I}_{B_0}$, $\mathcal{I}_{B_1}$, and $\mathcal{I}_\omega$ both away from and near the TPT.
		(d)-(f) QFIs $\mathcal{I}_{B_0}$, $\mathcal{I}_{B_1}$, and $\mathcal{I}_\omega$ versus the time $t$ over one time period for the fixed TPT condition ($B_0=B_1=0.5$). The standard quantum limit (SQL) $\sim t$, the Heisenberg limit (HL) $\sim t^2$, and the scaling of $\sim t^4$ are also exhibited for the comparison.}
	\label{Fig_I_groundstate}
\end{figure*}

In the following, we consider one time-periodic dynamics evolution as the example, which is widely investigated in the relevant Floquet sensing schemes~\cite{Lang2015,Ba2021}. 
The stroboscopic measurement time is set as $t=T_0=2\pi$ with $\omega_0=1$.
The Floquet indices $n,m=\{0,\pm 1, \cdots, \pm 50\}$ are set for the numerical simulations. 
The theoretical infinite-dimensional Floquet Hamiltonian (\ref{eq:floquet}) is truncated as the $202 \times 202$ matrix with $T=2\pi/\omega$ ($\omega=1$ is set for the simplification).
A convergence analysis of the QFI confirms that the chosen truncation size of the Floquet Hamiltonian is sufficient. Moreover, we use a finite-difference step size of $10^{-6}$ in the numerical simulations. The corresponding convergence and numerical-stability discussions are presented in appendix~C.
In addition, we choose the ground state $(|0\rangle - |1\rangle)/\sqrt{2}$ of $H(0)$ as the initial probe state,  which is the instantaneous ground state at $t=0$ and thus the most natural and experimentally relevant initialization. 
The QFIs $\mathcal{I}_{B_0}$, $\mathcal{I}_{B_1}$ and $\mathcal{I}_\omega$ for the parameters $B_0,B_1$, and $\omega$ are displayed in figures~\ref{Fig_I_groundstate} (a)-(c).  
The numerical simulations are performed using dimensionless parameters $B_0$, $B_1$, and $\omega$. The mapping between these parameters and the corresponding physical quantities is further discussed in appendix~C.
In figures~\ref{Fig_I_groundstate} (a)-(b), the QFIs increase sharply with oscillations around the TPT and reach the individual maximum at the TPT. 
The QFI oscillations away from and near the TPT are also plotted in the inset figures of figures~\ref{Fig_I_groundstate} (a)-(b).
These oscillations in the QFIs originate from coherent interference in the Floquet dynamics. As $B_0$ changes, the quasienergy splitting and the overlaps with different Floquet modes vary correspondingly, which leads to oscillatory changes in the generator fluctuations and hence in the QFIs. 
Figure~\ref{Fig_I_groundstate} (c) exhibits a prominent QFI enhancement near the TPT (see the inset) for the large $B_0$ and $B_1$.

Furthermore, by investigating the QFI scaling over one time period, we find that the QFIs $\mathcal{I}_{B_0}$ and $\mathcal{I}_{B_1}$ for parameters $B_0$ and $B_1$ are beyond the SQL $\sim t$ and attain the HL $\sim t^2$ at the TPT as shown in figures~\ref{Fig_I_groundstate} (d)-(e). 
The scaling analysis in figure~\ref{Fig_I_groundstate} (f) demonstrates that the QFI $\mathcal{I}_\omega$ for the parameter $\omega$ exceeds the SQL and HL, even achieving the scaling of $\sim t^4$ at the TPT. 
These scaling exponents at the TPT do not depend on the specific value of $B_0=B_1$, we leverage $B_0=B_1=0.5$ in figure~\ref{Fig_I_groundstate}(d)-(f) as the representative one.
Furthermore, the higher scaling of $\omega$ has a clear physical origin. The frequency estimation benefits from the extra time accumulation in the phase shift $\omega t$ of the Hamiltonian (\ref{eq:H}), while the field-strength parameters $B_0$ and $B_1$ do not.

The component-resolved QFI provides a means to investigate the origin of the aforementioned precision enhancement.
When far from the TPT but not particularly far away, the QFI is mainly contributed by the Floquet eigenmodes, i.e.,
\begin{eqnarray}\label{eq:QFI1}
	\mathcal{I}_{B_0} \simeq {I}_{B_0,\alpha},\quad
	\mathcal{I}_{B_1} \simeq {I}_{B_1,\alpha},\quad
	\mathcal{I}_{\omega} \simeq {I}_{\omega,\alpha}.
\end{eqnarray}
Near the TPT, the quasienergy gap gradually closes such that governs the QFI~\cite{Ba2021}. The hybridization among all components commonly gives the maixmal QFI.
For the parameters $B_0$ and $B_1$, no multi-photon process exists since $\partial_{\omega}/\partial B_0(B_1)=0$, and the resulting QFIs are 
\begin{align}\label{eq:QFI2}
	\mathcal{I}_{B_0}&=I_{{B_0},\alpha}+I_{{B_0},\varepsilon}+8\text{Cov}[\mathfrak{h}_{B_0,\alpha},\mathfrak{h}_{B_0,\varepsilon}]\;,\nonumber\\
	\mathcal{I}_{B_1}&=I_{{B_1},\alpha}+I_{{B_1},\varepsilon}+8\text{Cov}[\mathfrak{h}_{B_1,\alpha},\mathfrak{h}_{B_1,\varepsilon}]\;.
\end{align}
For the parameter $\omega$, its QFI also includes multi-photon processes and the relevant coherence terms, which gives
\begin{eqnarray}\label{eq:QFI3}
	\mathcal{I}_{\omega} &=& {I}_{\omega,\alpha}
	{+{I}_{\omega,\varepsilon}}
	{+{I}_{\omega,\omega}} \nonumber\\
	&+&8(\text{Cov}[\mathfrak{h}_{\omega,\alpha},\mathfrak{h}_{\omega,\varepsilon}]
	+\text{Cov}[\mathfrak{h}_{\omega,\alpha},\mathfrak{h}_{\omega,\omega}]+\text{Cov}[\mathfrak{h}_{\omega,\varepsilon},\mathfrak{h}_{\omega,\omega}]). \nonumber\\
\end{eqnarray}
Besides, our simulations can confirm that the discrepancy between the QFIs computed with and without the decomposition (corresponding to the first and second lines of equation~(\ref{eq:Ig}), respectively) is within $10^{-6}$.

\begin{figure*}[htp]
	\centering
	\includegraphics[width=0.8\textwidth]{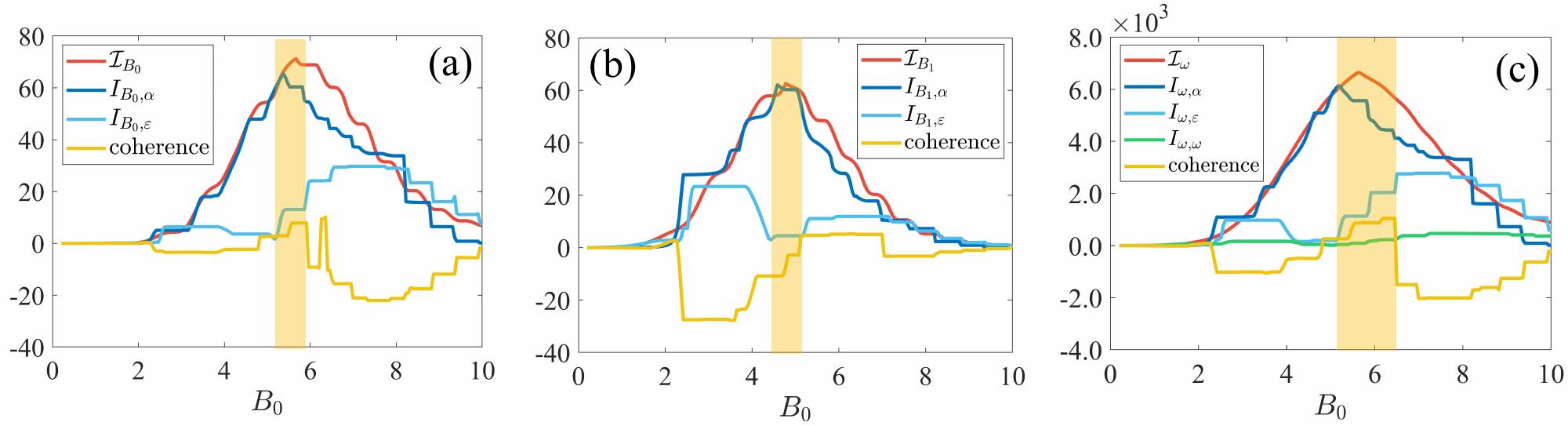}
	\caption
	{(a)-(c) Total QFI versus its Floquet-resolved components (eigenmodes, quasienergies, multi-photon processes, and coherence) for the parameters $B_0$, $B_1$, and $\omega$ with the fixed $B_1=5$. The TPT region around $B_0=B_1=5$ is indicated by the shaded area.}
	\label{Fig.resolve}
\end{figure*}

To illustrate equations~(\ref{eq:QFI1})–(\ref{eq:QFI3}), we plot the total QFI together with all its components for the parameters $B_0$, $B_1$ and $\omega$ in figures~\ref{Fig.resolve}(a)–(c), where the local mean extracted from the oscillatory signals~\cite{fitting} is used.
As shown in figures~\ref{Fig.resolve} (a)-(c), the QFI near the TPT (indicated by the shaded area) is governed by the interplay of all components, while away from the transition, contributions from the Floquet eigenmodes become dominant despite the other components remaining finite.
Moreover, the quasienergy-induced contribution (light-blue curves) exhibits an almost complementary behavior to that arising from the coherence among components (yellow curves) in figures~\ref{Fig.resolve}(a)–(c). 
The multi-photon processes provide no appreciable enhancement to the QFI of $\omega$, which is nearly zero (green curve) in figure~\ref{Fig.resolve}(c). 

\begin{figure*}[htp]
	\centering
	\includegraphics[width=0.85\textwidth]{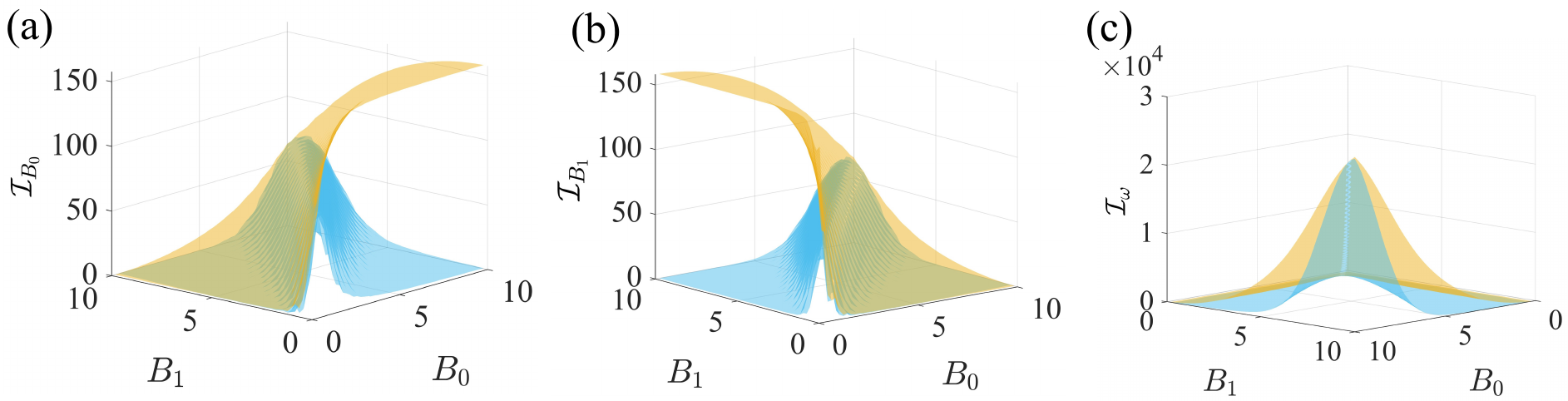}
	\caption
	{(a)-(c) Comparison of the QFI (blue surface) and its upper bound (yellow surface) for the parameters $B_0$, $B_1$, and $\omega$.  These two surfaces are overlapped at the TPT.}
	\label{HL}
\end{figure*}

Furthermore, we also focus on the connection of the QFI value the QFI upper bound in the vicinity of the TPT.
The comparisons of the QFI and its upper bound for parameters $B_0$, $B_1$, and $\omega$ are separately plotted in figures~\ref{HL} (a)-(c).
One can see that the QFI not only exhibits a peak, but more significantly, this QFI peak reaches the QFI upper bound at the TPT.
Figures~\ref{HL} (a)-(c) further imply that the metrological enhancement near the TPT is not solely tied to the specific choice of probe state. 
As a matter of fact, the saturation of the QFI upper bound typically requires preparing the large-scale entanglement or the high-quality squeezing in the experiments.
This way of driving the quantum system to the TPT provides an alternative pathway that mitigates these experimental challenges and reaches the ultimate precision limit.

\subsection{Incompatibility and stroboscopic projective measurement}
\begin{figure*}[htp]
	\centering
	\includegraphics[width=0.65\textwidth]{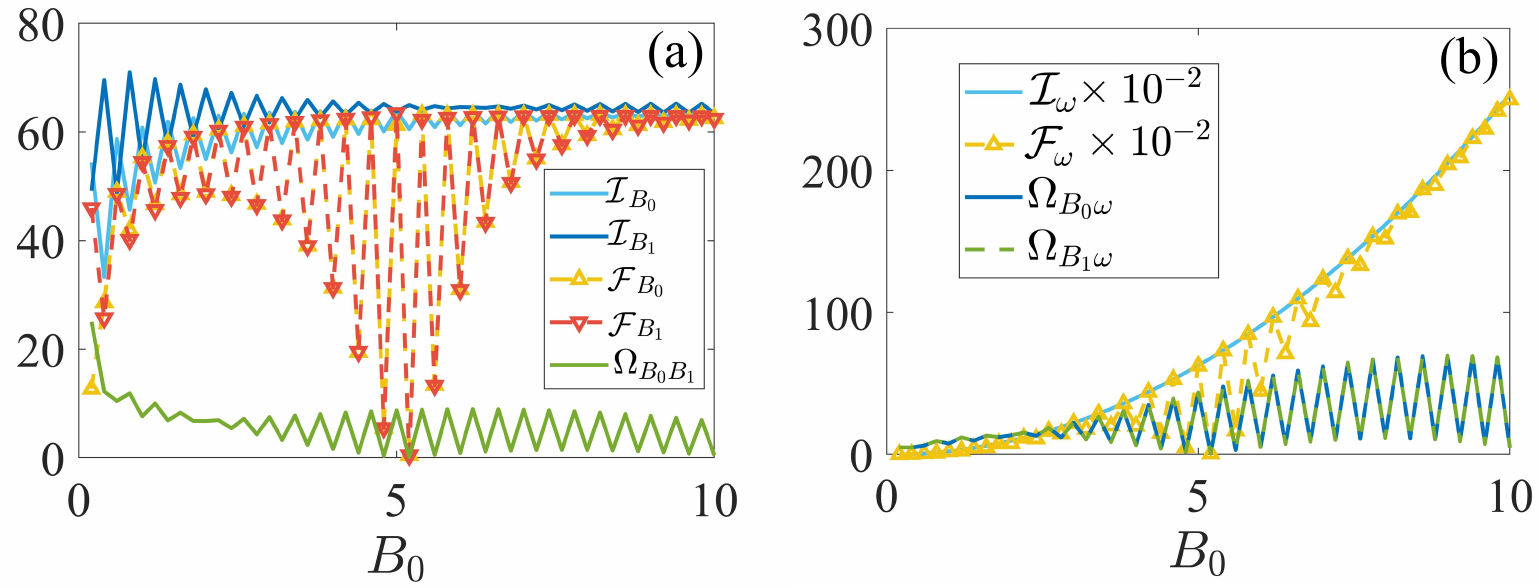}
	\caption
	{(a) Measurement incompatibility $\Omega_{B_0B_1}$ for the parameters $B_0$ and $B_1$ (b) $\Omega_{B_0 \omega}$ for the parameters $B_0$ and $\omega$, $\Omega_{B_1 \omega}$ for the parameters $B_1$ and $\omega$, and the corresponding QFIs $\mathcal{I}_{B_0}$, $\mathcal{I}_{B_1}$, and $\mathcal{I}_{\omega}$ along the TPT boundary.
		The classical Fisher informations (CFIs) $\mathcal{F}_{B_0}$, $\mathcal{F}_{B_1}$, and $\mathcal{F}_\omega$ obtained from the stroboscopic projective measurement at the end of one period are exhibited concurrently.}
	\label{MI}
\end{figure*}

Based on the weak commutation condition of equation~(\ref{eq:Mu}), the variations of measurement incompatibility among parameters $B_0$, $B_1$, and $\omega$ along the TPT boundary are depicted in figures~\ref{MI} (a)-(b). 
For the parameters $B_0$ and $B_1$, the incompatibility $\Omega_{B_0B_1}$ vanishes with oscillations, meanwhile, the QFIs $\mathcal{I}_{B_0}$, $\mathcal{I}_{B_1}$  gradually stabilize for large $B_0$. 
For the parameters $B_0 (B_1)$ and $\omega$, the incompatibility $\Omega_{B_0 \omega}$ ($\Omega_{B_1 \omega}$) oscillates around the zero value, and the QFI $\mathcal{I}_\omega$ increases with the enhancement of $B_0$. 
Accordingly, the simultaneous optimal estimation with respect to these three parameters can be guaranteed at the TPT for large $B_0$.
Physically, the suppression of the incompatibility at large $B_0$ can be understood from the generator perspective. Along the TPT boundary, when $B_0$ becomes large, the Floquet dynamics is dominated by the strong rotating-field term in Eq.~(30). The dominant parts of the generators with respect to different parameters become increasingly similar, while their differences remain only as subleading corrections. Therefore, the incompatibility induced by the noncommutativity between generators tends to vanish.
Additionally, in the experiments, the zero-incompatibility points can be identified by scanning the control parameter along the TPT boundary.
The experimenter does not need to know these points a priori but determines them from the measured variation trend of the incompatibility signal ($\Omega_{B_0 B_1},\Omega_{B_0 \omega}$ and $\Omega_{B_1 \omega}$) under a local parameter scan.
Since the incompatibility is continuously suppressed near these points, tuning sufficiently close to them may already be adequate for near-compatible multiparameter estimation.

Furthermore, projective measurements are typically employed in the NV center sensing experiments to extract unknown frequency information~\cite{QO,Lu2025,WangPRX}.
Accordingly, the projective measurement is considered as a candidate scheme within our Floquet framework.
For instance, we implement a stroboscopic projective measurement by projecting the state onto the excited state $|1\rangle$ at the end of each driving period. At the stroboscopic measurement time $t=lT_0$ for $l \in \mathbb{N}+$, the CFI form can be further simplified (see appendix~B).
Under the same simulation conditions utilized for the QFIs analyses, we can individually get the CFIs $\mathcal{F}_{B_0}$, $\mathcal{F}_{B_1}$, and $\mathcal{F}_\omega$ over one time period, which are plotted in figures~\ref{MI} (a)-(b).  
The CFIs quantitatively match the QFIs along the TPT boundary, as evidenced by their oscillatory coincidence in figures~\ref{MI}(a)-(b).
These characterizations of the CFIs originate from the competition among Floquet eigenmodes, quasienergies, multi-photon process, and the coherence of these three components as described in equation~(\ref{eq:Fell}). 
For large $B_0$, the curves of CFIs and QFIs for parameters $B_0$, $B_1$, and $\omega$ entirely overlap. 

\section{Rotating magnetic field sensing}\label{Sec:Rotate}
To further confirm the validity of the proposed Floquet framework, we also apply it to the paradigmatic problem of sensing a rotating magnetic field. The to-be-estimated parameters are the amplitude $B$ and the rotating frequency $\omega$ of the field.
A spin-1/2 particle interacts with a rotating magnetic field, which is described by the interaction Hamiltonian
\begin{eqnarray}\label{eq:Hrotate}
	H(t)=-B\left[\cos(\omega t)\sigma_x+\sin (\omega t)\sigma_z\right].
\end{eqnarray}
\begin{widetext}
By investigating the time evolution operator in the interaction picture, the generators for parameters $B$ and $\omega$ over one time period are inferred as~\cite{Pang2017,Hou2021}
\begin{align}
	h_B(T)&=\! \int_0^T dt U^\dagger(t) \left(\frac{\partial H(t)}{\partial B}\right) U(t)\nonumber\\
	&=\! -\left(\frac{4B^2 T}{4B^2+\omega^2}+\frac{\omega^2 \sin(T\sqrt{4B^2+\omega^2})}{(4B^2+\omega^2)^{3/2}} \right)\sigma_x\nonumber\\
	&+\! 2B\omega\left( \frac{T}{4B^2+\omega^2}-\frac{\sin(T\sqrt{4B^2+\omega^2})}{(4B^2+\omega^2)^{3/2}}\right)\sigma_y
	-\! \omega \! \left(\frac{1-\cos(T\sqrt{4B^2+\omega^2})}{4B^2+\omega^2}\right)\sigma_z\;,\label{eq:hB}\\
	h_\omega(T)&=\! \int_0^T dt U^\dagger(t) \left(\frac{\partial H(t)}{\partial \omega}\right) U(t)\nonumber\\
	&=\! B\left( \! \frac{\sin(T\sqrt{4B^2+\omega^2})}{(4B^2+\omega^2)^{3/2}} \!-\! \frac{T \cos(T\sqrt{4B^2+\omega^2})}{4B^2+\omega^2} \! \right)\! \nonumber\\
	&\times\! (\omega \sigma_x+2B \sigma_y)
	\!+\! B \! \left(\frac{-T\sin(T\sqrt{4B^2+\omega^2})}{\sqrt{4B^2+\omega^2}} 
	+\! \frac{1-\cos(T\sqrt{4B^2+\omega^2})}{4B^2+\omega^2} \right)\sigma_z\;.\label{eq:homega}
\end{align}
The analytical upper bounds on the QFI are derived from equation~(\ref{eq:IMax}) as
\begin{align}
	\mathcal{I}_{B}^M&=\frac{16B^2 T^2}{4B^2+\omega^2}+\frac{8\omega^2 [1-\cos(T\sqrt{4B^2+\omega^2})]}{(4B^2+\omega^2)^2},\label{eq:Ba}\\
	\mathcal{I}_\omega^M&=\frac{4B^2 T^2}{4B^2+\omega^2}-\frac{8B^2 T\sin(T \sqrt{4B^2+\omega^2})}{(4B^2+\omega^2)^{3/2}}
	+\frac{8B^2[1-\cos(T\sqrt{4B^2+\omega^2})]}{(4B^2+\omega^2)^2}.\label{eq:Bomega}
\end{align}
Given the probe state $(|0\rangle-|1\rangle)/\sqrt{2}$ (i.e., the ground state of $H(0)$), the weak commutation condition of equation~(\ref{eq:Mu}) leads to
\begin{eqnarray}\label{eq:weak}
	\Omega_{B\omega}=\frac{8B^2 T \omega [\cos(T\sqrt{4B^2+\omega^2})-1]}{(4B^2+\omega^2)^2}+\frac{4B^2 T^2 \omega \sin(T\sqrt{4B^2+\omega^2})}{(4B^2+\omega^2)^{3/2}}\;.
\end{eqnarray}

Based on the Floquet theory of section~\ref{Sec:Floquet}, we derive the Floquet Hamiltonian in terms of (\ref{eq:Hrotate}) as
\begin{eqnarray}
	H_F=\left(\begin{array}{cccccccccccc}
		\ddots&\vdots&\vdots&\vdots&\vdots&\vdots&\vdots&\vdots&\vdots&\vdots&\vdots&\begin{sideways}
			$\ddots$
		\end{sideways}\\
		\cdots& -2\omega & 0 & -\frac{i B_0}{2} & -\frac{{B_0}}{2} & 0 & 0 & 0 & 0 & 0 & 0 &\cdots \\
		\cdots& 0 & -2\omega & -\frac{{B_0}}{2} & \frac{i B_0}{2} & 0 & 0 & 0 & 0 & 0 & 0 &\cdots\\
		\cdots&	\frac{i B_0}{2} & -\frac{{B_0}}{2} & -\omega & 0 & -\frac{{i B_0}}{2} & -\frac{{B_0}}{2} & 0 & 0 & 0 & 0 &\cdots\\
		\cdots&	-\frac{{B_0}}{2} & -\frac{{i B_0}}{2} & 0 & -\omega & -\frac{{B_0}}{2} & \frac{i B_0}{2} & 0 & 0 & 0 & 0 &\cdots\\
		\cdots&	0 & 0 & \frac{i B_0}{2} & -\frac{{B_0}}{2} & 0 & 0 & -\frac{{i B_0}}{2} & -\frac{{B_0}}{2} & 0 & 0 &\cdots\\
		\cdots&	0 & 0 & -\frac{{B_0}}{2} & -\frac{{i B_0}}{2} & 0 & 0 & -\frac{{B_0}}{2} & \frac{i B_0}{2} & 0 & 0 &\cdots\\
		\cdots&	0 & 0 & 0 & 0 & \frac{i B_0}{2} & -\frac{{B_0}}{2} & \omega & 0 & -\frac{{i B_0}}{2} & -\frac{{B_0}}{2} &\cdots\\
		\cdots&	0 & 0 & 0 & 0 & -\frac{{B_0}}{2} & -\frac{{i B_0}}{2} & 0 & \omega & -\frac{{B_0}}{2} & \frac{i B_0}{2} &\cdots\\
		\cdots&	0 & 0 & 0 & 0 & 0 & 0 & \frac{i B_0}{2} & -\frac{{B_0}}{2} & 2\omega & 0 &\cdots\\
		\cdots&	0 & 0 & 0 & 0 & 0 & 0 & -\frac{{B_0}}{2} & -\frac{{i B_0}}{2} & 0 & 2\omega&\cdots\\
		\begin{sideways}
			$\ddots$
		\end{sideways}&\vdots&\vdots&\vdots&\vdots&\vdots&\vdots&\vdots&\vdots&\vdots&\vdots&\ddots
	\end{array}\! \right).
\end{eqnarray}

We then employ the Floquet framework presented in section~\ref{Subsec:Fparameter} to numerically compute the QFI upper bounds and the weak commutation condition for parameters $B$ and $\omega$.
The probe state remains the ground state of $H(0)$, namely $(|{0}\rangle-|{1}\rangle)/\sqrt{2}$. 
As shown in figures~\ref{rotate} (a)-(b), the QFI maximum for the parameter $B$ scales as $\sim t^2$ (HL), and the scaling reaches $\sim t^4$ for the parameter $\omega$. 
Although the scaling $\sim t^4$ for the frequency estimation appears in both the Rashba model and the rotating magnetic-field model, this is not a universal property of all periodically driven two-level systems within the Floquet framework. 
Whether this scaling can be realized depends on the specific form of the time-periodic Hamiltonian, such as the driving protocol and the working point in parameter space, especially whether the QFI can approach its upper bound.
The Rashba model and the rotating magnetic field model are two representative examples exhibiting this common mechanism.

Furthermore, the characterizations of the QFIs and measurement incompatibility for this state are also investigated.
The weak commutation condition between $B$ and $\omega$ vanishes with the increase of $B$ as demonstrated in figure~\ref{rotate} (c).
The numerical results are in excellent agreement with the analytical expressions of equations~(\ref{eq:Ba})-(\ref{eq:weak}), which benchmarks the validity of our Floquet approach.
\begin{figure}[!h]
	\centering
	\includegraphics[width=0.85\textwidth]{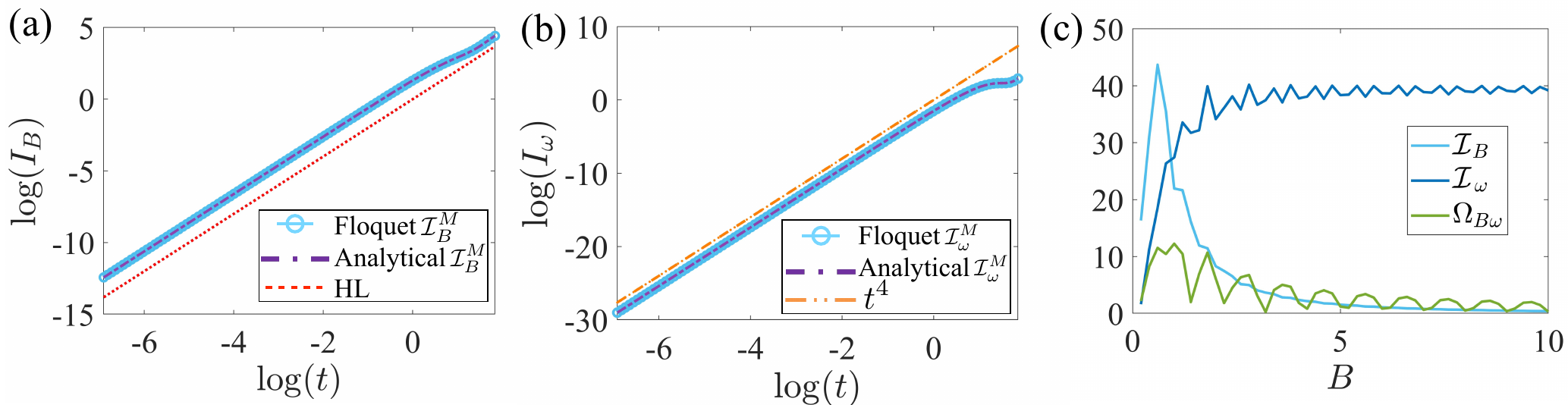}
	\caption
	{Comparison among the QFI upper bounds for the parameters $B$ and $\omega$ obtained from the proposed Floquet approach, the analytical results of equations~(\ref{eq:Ba})-(\ref{eq:Bomega}), and (a) the HL $\sim t^2$ or (b) the scaling of $\sim t^4$, over one time period for the fixed $B=0.5$. 
		(c) The Floquet-derived QFIs $\mathcal{I}_B$, $\mathcal{I}_\omega$, and the measurement incompatibility $\Omega_{B\omega}$ as functions of the parameter $B$.}
	\label{rotate}
\end{figure}
\end{widetext}

\section{Conclusions and discussions}\label{Sec:Conclusion}
We have concentrated on the quantum multiparameter estimation problems in generic time-periodically driven systems.
A quantum multiparameter estimation protocol in the Floquet theory framework has been established, in which the estimation precision contributed from Floquet eigenmodes, quasienergies, and multi-photon processes has been determined, respectively.
Moreover, we have applied the proposed Floquet approach in a ring-shaped Rashba spin-orbit interferometer with the periodically driven-induced TPT.
By driving the system to the TPT boundary, the estimation precision for the magnetic field strengths can be dramatically enhanced and reach the HL $\sim t^2$, and that scales to the higher $t^4$ for the frequency estimation.
The measurement incompatibility can be eliminated in the fashion of oscillations, and the stroboscopic projective measurement makes the QCRB achievable.
Furthermore, an investigation of multiparameter estimation in a rotating magnetic field has provided a further benchmark for the validity of our approach.

Our Floquet multiparameter estimation strategy provides a possible pathway to achieve the high precision estimation by modulating the distinct QFI components. 
Specifically, one can only select the Floquet sidebands that carry the highest parameter sensitivity and discard the remaining sidebands, which is exactly the theme of utilizing Floquet engineering in quantum sensing~\cite{An2023}.
Besides, how to generalize this approach to more complex settings, for instance, the many-level quantum system subject to the polychromatic field~\cite{MF,MF2010,WangPRX}, will be our next task.

\begin{acknowledgments}
Yuyang Tang has the same contributions for this work with Yu Yang.
We thanks for the useful discussions with Gaoxiang Li from CCNU, Haidong Yuan from CHUK, and Di Zhao from XJTU. 
This research is supported by the Fundamental Research Funds for the Central Universities (Grant No. xxj032025044),
Shaanxi Fundamental Science Research Project for Mathematics and Physics (Grants No. 23JSQ012),
National Natural Science Foundation of China (NSFC) (Grants No. 12474363).
\end{acknowledgments}

\section*{Appendix A. Spin texture in the Frenet-Serret frame}\label{App:Spintexture}
In the Frenet–Serret frame determined by the tangential ($\vec{T}$), normal ($\vec{N}$), and binormal ($\vec{B}$) directions, the Pauli matrices $\hat{\sigma}=\{{\sigma}_x,{\sigma}_y,{\sigma}_z\}$ can be renewed as $\hat{\sigma}_\text{FS}=\{\sigma_T(t),\sigma_N(t),\sigma_B\}$ with 
\begin{eqnarray}
	&&\sigma_T(t)= \vec{T}\cdot\hat{\sigma}=\sin \vartheta(t) \sigma_x-\cos \vartheta(t) \sigma_y\;,\label{eq:sigmaT}\\
	&&\sigma_N(t)= \vec{N}\cdot\hat{\sigma}=\cos \vartheta(t)\sigma_x+\sin \vartheta(t) \sigma_y\;,\\
	&&\sigma_B=\vec{B}\cdot\hat{\sigma}=\sigma_z\;,
\end{eqnarray}
where $\vartheta(t)$ is defined by equation~(\ref{eq:Bt}) in the maintext.
We further get $\partial_t \vec{N}=K(t) \vec{T}$ and $\partial_t \vec{T}=-K(t) \vec{N}$ regarding the curvature of equation~(\ref{eq:curvature}).
Consequently, the Hamiltonian of equation~(\ref{eq:genH}) is rewritten into
\begin{eqnarray}\label{eq:Ht}
	H(t)=\mathcal{B}(t)\sigma_N(t)\;.
\end{eqnarray}
The local spin orientation for a given spin eigenmode is defined as the expectation value of the spin operators in the Frenet–Serret frame, i.e.,
$\langle \hat{\sigma}_\text{FS} \rangle=\{ \langle \sigma_T(t)\rangle, \langle \sigma_N(t)\rangle, \langle \sigma_B \rangle\}$.
Based on the Schr\"odinger equation for the spin eigenmode, one has
\begin{eqnarray}\label{eq:Pspin}
	\partial_t \langle \hat{\sigma}_\text{FS}\rangle
	\!=\!\frac{i}{\hbar}\langle [H(t),\hat{\sigma}_\text{FS}]\rangle\!+\!\langle  \partial_t \hat{\sigma}_\text{FS}\rangle, 
\end{eqnarray} 
which describes the evolution of the local spin orientation in time.
According to equations~(\ref{eq:sigmaT})-(\ref{eq:Ht}), we further obtain $[H(t),\sigma_N(t)]=0$, $[H(t),\sigma_T(t)]=-2i \mathcal{B}(t) \sigma_z$, $[H(t),\sigma_z]=2i\mathcal{B}(t)\sigma_T(t)$, 
$\partial_t \langle \sigma_N(t)\rangle=K(t)\langle \sigma_T(t)\rangle$, 
$\partial_t \langle \sigma_T(t)\rangle=\frac{2}{\hbar}\mathcal{B}(t) \langle \sigma_z\rangle -K(t)\langle \sigma_N(t)\rangle$ and $\partial_t \langle \sigma_z\rangle=\frac{-2}{\hbar} \mathcal{B}(t) \langle \sigma_T(t)\rangle$.
Equation~(\ref{eq:Pspin})  therefore can be expressed by a {\it gyroscope-like form}~\cite{Book2002}
\begin{eqnarray}\label{eq:spin}
	\partial_t \langle \hat{\sigma}_\text{FS}\rangle=\hat{H}_\text{eff}(t) \times \langle \hat{\sigma}_\text{FS} \rangle\;,
\end{eqnarray}
with an effective spin-orbit field
\begin{eqnarray}
	\hat{H}_\text{eff}(t)=\left\{0,\frac{2 \mathcal{B}(t)}{\hbar},K(t)\right\}\;.
\end{eqnarray}

\section*{Appendix B. Stroboscopic detection}\label{App:SM}
At the stroboscopic detection time $t=lT_0=2l\pi/\omega_0$ for $l \in \mathbb{N}+$ with $\omega_0$ being the external clock frequency, the generator's elements of equation~(\ref{eq:ele}) simplify into
\begin{eqnarray}
	&&{\hspace{-0.5cm}}\langle \gamma|\mathfrak{h}_{\ell,\alpha}(lT_0)|\beta\rangle= i \sum_{\mu,\alpha,k} \frac{\partial \widetilde{B}_{\alpha k}}{\partial x_\ell} e^{-i\lambda_\alpha lT_0}   D_{\mu \beta}^*(lT_0),
	\nonumber\\
	&&{\hspace{-0.5cm}}\langle \gamma|\mathfrak{h}_{\ell,\varepsilon}(lT_0)|\beta\rangle=lT_0 \sum_{\mu,\alpha,k} \widetilde{B}_{\alpha k}\frac{\partial \lambda_\alpha}{\partial x_\ell} e^{-i\lambda_\alpha lT_0}  D_{\mu \beta}^*(lT_0),  \nonumber\\
	&&{\hspace{-0.5cm}}\langle \gamma|\mathfrak{h}_{\ell,\omega}(lT_0)|\beta\rangle \!=\! -k l T_0 \! \sum_{\mu,\alpha,k} \! \widetilde{B}_{\alpha k} \! \frac{\partial \omega}{\partial x_\ell} \! e^{-i\lambda_\alpha l T_0} \! D_{\mu \beta}^*\!(lT_0)\!.
\end{eqnarray}
The resulting QFI components are renewed as
\begin{align}
	{I}_{\ell,\alpha}&=4( \text{Tr}\left[\mathfrak{h}_{\ell,\alpha}^2(lT_0) \rho_0\right]-\text{Tr}^2\left[\mathfrak{h}_{\ell,\alpha}(lT_0) \rho_0\right] )\;,\nonumber\\
	{I}_{\ell,\varepsilon}&=4( \text{Tr}\left[\mathfrak{h}_{\ell,\varepsilon}^2(lT_0) \rho_0\right]-\text{Tr}^2\left[\mathfrak{h}_{\ell,\varepsilon}(lT_0) \rho_0\right] )\;,\nonumber\\
	{I}_{\ell,\omega}&=4 ( \text{Tr}\left[\mathfrak{h}_{\ell,\omega}^2(lT_0) \rho_0\right]-\text{Tr}^2\left[\mathfrak{h}_{\ell,\omega}(lT_0) \rho_0\right] ).
\end{align} 
Moreover, the CFI of equation~(\ref{eq:Fell}) simplifies into
\begin{widetext}
\begin{eqnarray}\label{eq:CFInew}
	\mathcal{F}_\ell = \sum_\gamma \! \frac{\! 4\text{Re}^2 \! \Big[D_{\beta\gamma}^*(lT_0)  \sum_{\alpha,k} \Big(
		{ \frac{\partial B_{\alpha k}}{\partial x_\ell}}
		{- i lT_0 B_{\alpha k} \frac{\partial \lambda_\alpha}{\partial x_\ell}}
		{+ik lT_0 B_{\alpha k} \frac{\partial \omega}{\partial x_\ell}}
		\Big)  e^{-i\lambda_\alpha l T_0}  \Big]}{|D_{\beta \gamma} ({lT_0})|^2}.
\end{eqnarray}
\end{widetext}

\section*{Appendix C. Numerical simulation details}\label{App:TR}
\begin{figure}[!h]
	\centering
	\includegraphics[width=0.3\textwidth]{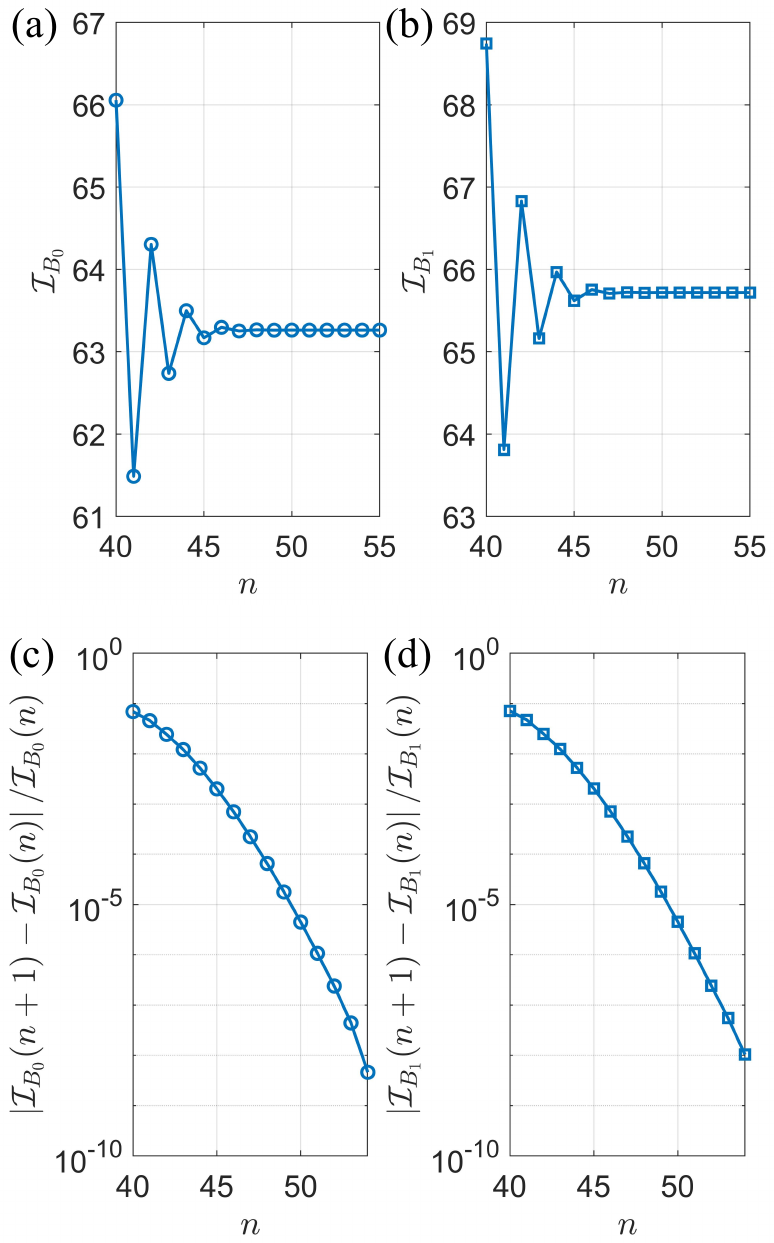}
	\caption
	{(a)-(b) Convergence dependence of QFIs $\mathcal{I}_{B_0}$ and $\mathcal{I}_{B_1}$ upon the Fourier indices $n$ for the fixed TPT condition ($B_0=B_1=10$).
		(c)-(d)  Relative change of QFIs $\mathcal{I}_{B_0}$ and $\mathcal{I}_{B_1}$ induced by increasing the cutoff from $n$ to $n+1$ for the fixed TPT condition ($B_0=B_1=10$).}
	\label{conve}
\end{figure}

Here we explicitly demonstrate the convergence of the QFIs with respect to the truncation size of the Floquet Hamiltonian near the TPT. Since the truncation is determined by the range of Fourier indices, we examine how the QFIs $\mathcal{I}_{B_0}$ and $\mathcal{I}_{B_1}$ vary as the cutoff $n$ is increased, as plotted in figure~{\ref{conve}}(a)-(d).
Specifically, figure~{\ref{conve}}(a)-(b) show that both $\mathcal{I}_{B_0}$ and $\mathcal{I}_{B_1}$ have already converged for $n>46$. To quantify the convergence more clearly, we further define the relative change induced by increasing the cutoff from $n$ to $n+1$ as
$\frac{|\mathcal{I}_{\ell}(n+1)-\mathcal{I}_{\ell}(n)|}{\mathcal{I}_{\ell}(n)}$ for $x_\ell \in \{B_0,B_1\}$, and plot it in figure~\ref{conve}(c)-(d). The results show that this relative difference falls below $10^{-5}$ at $n=50$ for both QFIs.
Therefore, truncating the Floquet Hamiltonian with Fourier indices up to $\pm 50$ (corresponding to a $202\times 202$ matrix in our two-level model) is sufficient to ensure that the reported QFI results are numerically converged. 

\begin{figure}[!h]
	\centering
	\includegraphics[width=0.4\textwidth]{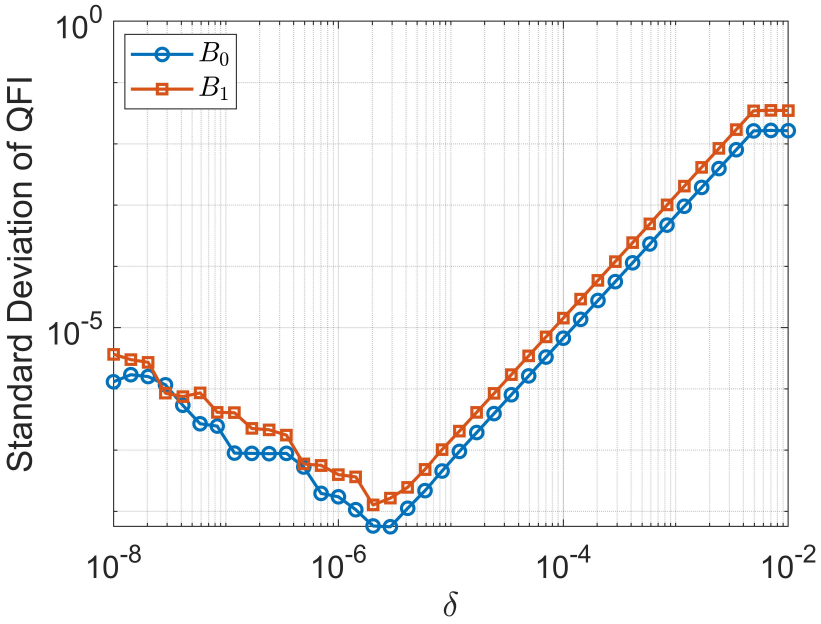}
	\caption
	{QFIs $\mathcal{I}_{B_0}$ and $\mathcal{I}_{B_1}$ versus the finite-difference step size $\delta$ for the fixed TPT condition ($B_0=B_1=0.5$), with the horizontal axis on a logarithmic scale.}
	\label{step}
\end{figure}

In addition, the step size in the center difference method is crucial for the numerical evaluation of the generator and the QFI, we use $\delta=10^{-6}$ in our simulations as the step size. 
To quantify the numerical stability, we evaluate the local standard deviation of the QFI using five adjacent step sizes around each step size $\delta_i$.
The dependence of the QFI on the step size near the TPT is plotted in figure~\ref{step}.
Since the generator is computed from the finite-difference formula $\frac{\partial U(t)}{\partial x_\ell}\approx \frac{U(x_\ell+\delta_i,t)-U(x_\ell-\delta_i,t)}{2\delta_i}$ for $x_\ell=\{B_0,B_1\}$,
the step-size dependence of the QFI directly reflects the numerical accuracy of this approximation. 
Figure~\ref{step} shows that both overly large and overly small step sizes lead to fluctuations in $\mathcal{I}_{B_0}$ and $\mathcal{I}_{B_1}$, whereas the fluctuations are minimized around $\delta=10^{-6}$. We therefore conclude that $\delta=10^{-6}$ is an appropriate choice for the simulations reported in this work.

Besides, we also further clarify the mapping between the dimensionless parameters in the Hamiltonian (\ref{eq:H}) and the corresponding physical quantities. 
Introducing a reference energy scale $E_0$, the dimensionless parameters are related to the physical ones by
\begin{eqnarray}
	B_0=\frac{g^*\mu_B B_{\rm ac}}{2E_0},\quad
	B_1=\frac{g^*\mu_B B_{\rm dc}}{2E_0},\quad
	\omega=\frac{\hbar\Omega}{E_0},
\end{eqnarray}
where $g^*$ is the effective electron $g$-factor, $\mu_B$ is the Bohr magneton, $\Omega$ is the driving angular frequency, and $B_{\rm ac}$ and $B_{\rm dc}$ are the physical magnetic-field amplitudes. 
For example, we set $E_0=\hbar\Omega$ i.e., measuring the energies in units of the drive frequency. In this case, $\omega=1$ and $B_0=B_1=0.5$ imply $B_{\rm ac}=B_{\rm dc}=\frac{\hbar\Omega}{|g^*|\mu_B}=\frac{f}{|g^*|(\mu_B/h)}$  with $\Omega=2\pi f$ and $\hbar=h/2\pi$. Using $\mu_B/h\simeq 13.996  \mathrm{GHz/T}$, one obtains $B_{\rm ac}=B_{\rm dc}\simeq \frac{f}{13.996\,|g^*|}\ \mathrm{T}$.
For an InP/InGaAs/InAlAs quantum well, the effective electron $g$-factor is material-dependent. 
For instance, we take a representative value $|g^*|\simeq 4$ presented in~\cite{gfactor}, then one further gets $B_{\rm ac}= B_{\rm dc}\simeq \{0.18\text{T},0.36\text{T},1.07\text{T}\}$ for $f=\{10\text{GHz},20\text{GHZ},60\text{GHZ}\}$.
Therefore, the simulated operating point corresponds to experimentally reasonable microwave-frequency driving (GHz to tens of GHz) and magnetic fields ($0.1$ to $1$ Tesla).


\begin{thebibliography}{99}
\bibitem{Explain}
	The Fourier factor satisfies $\langle t|m\rangle=e^{i m \omega t}$ for the time vector $|t\rangle$, which indicates that the probe state is $|\beta,0\rangle$, since $\langle 0|m\rangle=1$ at the initial time $t=0$.
\bibitem{fitting}
	The midline extraction removes fast oscillations and allows us to focus on the background behavior near the TPT. Other extraction procedures, such as fitting the peaks with a Lorentzian lineshape on top of a quadratic polynomial background, are also applicable, but are not essential to our proposal and are used only to extract the slowly varying trend.
\end{thebibliography}
\end{document}